# The Two-Component Quantum Theory of Atoms in Molecules (TC-QTAIM): The unified theory of localization/delocalization of electrons, nuclei and exotic elementary particles

Mohammad Goli and Shant Shahbazian[*]

*Faculty of Chemistry, Shahid Beheshti University, G. C. , Evin, Tehran, Iran, 19839, P.O. Box 19395-4716.*

Tel/Fax: 98-21-22431661

E-mail:
(Shant Shahbazian) chemist_shant@yahoo.com

[*] Corresponding author

## Abstract

In this contribution, pursuing our research program extending the atoms in molecules analysis into unorthodox domains, another key ingredient of the two-component quantum theory of atoms in molecules (TC-QTAIM) namely, the theory of localization/delocalization of quantum particles, is disclosed. The unified proposed scheme is able not only to deal with the localization/delocalization of electrons in/between atomic basins, but also to treat nuclei as well as exotic particles like positrons and muons equally. Based on the general reduced second order density matrices for indistinguishable quantum particles, the quantum fluctuations of atomic basins are introduced and then used as a gauge to quantify the localization/delocalization introducing proper indexes. The explicit mass-dependence of the proposed indexes is demonstrated and it is shown that a single localization/delocalization index is capable of being used for all kind of quantum particles regardless of their masses or charge content. For various non-Born-Oppenhiemer (non-BO) wavefunctions, including Hartree-product as well as singlet and triplet determinants, the indices are calculated and then employed to rationalize the localization/delocalization of particles in a series of four-body model systems consist of two electrons and two positively charged particles with variable mass. The ab initio FV-MC_MO derived non-BO wavefunctions for the four-body series are used for a comprehensive computational TC-QTAIM analysis, including topological analysis as well as basin integrations, in a wide mass regime, $m = 10 m_e - 10^{13} m_e$ ($m_e$ stands for electron mass), disclosing various traits in these series of species that are unique to the TC-QTAIM. On the other hand, it is demonstrated that in the large mass extreme the TC-QTAIM analysis reduces to the one performed within context of the orthodox QTAIM with two clamped positive particles revealing the fact that the TC-QTAIM encompasses the orthodox QTAIM as an asymptote. Finally, it is concluded that the proposed localization/delocalization scheme is capable of quantifying quantum tunneling of nuclei for systems containing delocalized protons. Such capability promises novel applications for the TC-QTAIM as well as its extended multi-component version (MC-QTAIM) introduced recently.

# 1 Introduction

Real space description of quantum systems has been always a matter of concern in quantum chemistry. This is understandable since *chemical discourse*, by its very nature and intricate historical development [1], is intrinsically tied with the *structural theory* seeking for 3D description of molecules and their higher-order aggregations. Accordingly, if one aims to go beyond the physical observables, seeking for *chemical observables*, e.g. atoms in molecules, bonding patterns, reactivity modes, etc. transforming the entities genuinely residing in Hilbert/configuration space to their counterparts in real space is inevitable [2,3]. This is not a trivial task since there is no general protocol for such transformation, a map from Hilbert to real space, in formal quantum mechanics [4]. Accordingly, *creative* methodologies have been developed for building bridges between the structural theory and quantum mechanics as seemingly desperate categorizers [3] where the quantum theory of atoms in molecules (QTAIM) is an illustrative [5-7], though not exclusive, example.

Recently, a generalized formalism has been developed extending the applicability domain of the orthodox QTAIM [8-11]. The extended methodology, termed the two-component quantum theory of atoms in molecules (TC-QTAIM), is competent to unravel the atoms in molecules from ab initio wavefunctions derived beyond the Born-Oppenhiemer (BO) paradigm as well as those of exotic species, e.g. wavefunctions of the positronic and muonic molecules [9,12-14]. On the other hand, it was demonstrated that the TC-QTAIM encompasses the orthodox, single component, QTAIM thus unifying the AIM description of various usual and unusual many-body quantum systems. It is particularly important to realize that in contrast to the orthodox QTAIM, the mass ratio of constituent bodies is a variable within the mathematical fabric of the TC-QTAIM directly affecting the morphology of boundaries as well as properties of the topological atoms. This is best illustrated in the case of isotopic nuclei since when treated as quantum waves, the TC-QTAIM derived topological atoms containing proton or deuterium (or tritium) nuclei are clearly distinguishable [8]. The capability of discerning topological atoms containing isotopic nuclei within the framework of the TC-QTAIM is of immediate importance in considering various isotope effects within the AIM methodology [15]. However, it must be noted that in contrast to its achievements, the TC-QTAIM is just the first step in the project of extending the orthodox QTAIM [11,16].

Based on this background, in this contribution a computational survey is done on the four-body species that are contained from two electrons and two identical quantum particles each carrying a unit of positive charge. This class of few-body systems includes the isotopomers of hydrogen molecule, muonium dimmer ($\mu_2$) as well as a series of bi-excitons; the latter had a long history with various applications in the solid state physics [17-26]. The real space structure of few-body exotic quantum systems has recently captured some attention [27-36] and is a particularly attractive field for the implementation of the TC-QTAIM analysis. Accordingly, the mass-dependence of the TC-QTAIM analysis is unraveled in this series of species in the whole considered mass spectrum, $m = 10 m_e - 10^{13} m_e$ ($m_e$ stands for electron mass), which mimics in the two extremes the clamped nuclei hydrogen molecule and almost a positronium dimer. Whereas all the previously developed theoretical tools are used in this computational consideration [8-10], a new theoretical entity namely the localization/delocalization analysis of the positively charged quantum particles within/between atomic basins is also introduced in this contribution. As is demonstrated in forthcoming discussions, the localization/delocalization of quantum particles explicitly depends on their mass revealing another non-trivial feature of the TC-QTAIM analysis.

Accordingly, the main goal of present study is supplementing previous contributions introducing the mathematical formalism of the localization/delocalization analysis within context of the TC-QTAIM including not only electrons but also quantum nuclei as well as exotic particles such as positrons and muons. The emerging *unified* scheme is capable of treating quantum systems beyond the BO paradigm thus revealing the *fine structure* of quantum particles' distribution within/between atomic basins which do not contain any clamped nucleus. It is particularly demonstrated that upon approaching the infinite mass limit of nuclei, where the clamped nucleus model is dominating, the unified scheme reduces to the usual localization/delocalization methodology, solely making an allowance for electrons, introduced previously within context of the orthodox QTAIM [37-44]. Like previous studies [8-10], this observation strengthens the claim that the orthodox QTAIM is just the *asymptotic* formulation of the TC-QTAIM or in other words, the latter *encompasses* the former.

The paper is organized as follows: First, the main cornerstones of the TC-QTAIM are briefly reviewed with some emphasis on the topological atoms and their properties. Then, in subsequent section the theoretical apparatus for the localization/delocalization analysis are disclosed introducing generalized localization/delocalization indexes. Particularly, analytical equations are introduced in this section revealing the explicit mass dependence of the localization/delocalization indexes. In the last section, computational TC-QTAIM analysis is done using the ab intio wavefunctions of the series of the four-body model systems described previously.

## 2 Basic principles of the TC-QTAIM

The theoretical framework of the TC-QTAIM has been disclosed very recently in full details [8-10] therefore, only a brief survey is done before introducing the ingredients of the localization/delocalization analysis. The TC-QTAIM is basically a complex of three cornerstones that have been offered as three independent axioms that all useful theorems as well as basic concepts and entities are derivable from these axioms [9].

The *axiom of form* (Axiom I) introduces the basic 3D scalar field, the Gamma field, used to decipher the atomic basins [9,11,16]:

$$\Gamma^{(2)}(\vec{q}) = \rho_-(\vec{q}) + (m_-/m_+)\rho_+(\vec{q}) \tag{1}$$

In this equation $m_-/m_+$ is the mass of electrons/positively charged quantum particles and $\rho_\pm(\vec{q}) = N_\pm \int d\tau'_\pm \Psi^*\Psi$ where $\int d\tau'_\pm$ denotes the sum over spin variables of all quantum particles and integration over all spatial coordinates except a single electron/the positively charged particle (PCP) designated through the subscript. Also, $N_-/N_+$ stands for the number of electrons/PCPs while $\rho_-(\vec{q})/\rho_+(\vec{q})$ acts as quantum one-densities of electrons/the PCPs that replace the usual one-electron density employed within context of the orthodox QTAIM [5-7]. The usual topological analysis, devised originally within the orthodox QTAIM, is also used to derive the topological atoms as the *basins of attraction* of the gradient vector field of the Gamma, $\vec{\nabla}\Gamma^{(2)}(\vec{q})$, within context of the TC-QTAIM.

The *axiom of choice* (Axiom II) is a recipe introducing the construction scheme of various *property densities* that after regional integration within each atomic basin yields the corresponding properties of topological atoms [9]:

$$\tilde{M}(\Omega) = \int_\Omega d\vec{q}\ \tilde{M}(\vec{q}) = \int_\Omega d\vec{q}\ \{M_-(\vec{q}) + M_+(\vec{q})\} = M_-(\Omega) + M_+(\Omega) \tag{2}$$

In this equation $M_-(\vec{q})/M_+(\vec{q})$ is property densities associated with electrons/the PCPs combined to yield the total density, $\tilde{M}(\vec{q})$, thus the target property, $\tilde{M}(\Omega)$, is the sum of contributions emerging from both electrons, $M_-(\Omega)$, and the PCPs, $M_+(\Omega)$. The sum of property contributions of all topological atoms of a molecule yields the total property attributed to the molecule, $M_{total} = \sum_\Omega \tilde{M}(\Omega)$.

The *axiom of rigidity* (Axiom III) is the last axiom that acts as a recipe for the construction of mechanical properties. This axiom *uniquely* determines the mechanical property densities through the extended local hypervirial theorem by introducing, albeit indirectly, the Schrödinger-Pauli-Epstein stress tensor density thus circumventing the problem of the non-uniqueness of the stress tensor densities [9,45].

$$\tilde{M}(\vec{q}) = \vec{\nabla} \bullet \tilde{J}_{\hat{G}}(\vec{q}) \tag{3}$$

In this equation $\hat{G} = \sum_i g_i$ stands for the sum of one-particle generators of a mechanical property while $\tilde{M}(\vec{q}) = M_-(\vec{q}) + M_+(\vec{q})$ and $\tilde{J}_{\hat{G}}(\vec{q}) = \vec{J}_{\hat{G}}^-(\vec{q}) + \vec{J}_{\hat{G}}^+(\vec{q})$. Accordingly, $M_\pm(\vec{q}) = (iN_\pm/\hbar) \int d\tau'_\pm \Psi^* ([\hat{H}, \hat{g}_\pm]) \Psi$ and $\hat{H}$ stand for mechanical property densities and the Hamiltonian, respectively, while $\vec{J}_{\hat{G}}^\pm(\vec{q}) = (N_\pm \hbar / 2m_\pm i) \int d\tau'_\pm \{\Psi^* \vec{\nabla}_q (\hat{g}_\pm \Psi) - (\hat{g}_\pm \Psi) (\vec{\nabla}_q \Psi^*)\}$ are the current property densities. Upon integration over the basins of topological atoms, described by equation (2), the regional contribution of the mechanical properties is derived. Based on the extended local hypervirial theorem, equation (3), various local theorems are derivable enabling one to compute mechanical properties of two-component systems [10]. The extended force, virial, torque, power, current and continuity theorems are just examples that have their counterparts within context of the orthodox QTAIM [46].

In the following section the strategy of constructing densities introduced by the axiom of choice is extended to include two-particle densities enabling one to develop extended localization/delocalization indexes.

## 3 Localization/delocalization of quantum particles: Foundations

The underpinnings of the electron localization/delocalization analysis in/between atomic basins had been worked out in detail by Bader and coworkers within context of the orthodox QTAIM [37-44]. While the orthodox analysis is confined to electrons as the sole quantum particles in one-component systems, in two-component systems one is faced basically with two kinds of localization/delocalization for each type of quantum particles. Consequently, the orthodox electron localization/delocalization analysis must be extended properly. In the following discussion this extended framework is considered briefly employing the same notation used in the original literature leaving a more comprehensive analysis to a future contribution.

### 3.1 The basic formalism

Assuming a quantum system composed of $N_-/N_+$ quantum particles, each located by a vector $\vec{r}_i^-/\vec{r}_i^+$, and partitioned into topological atoms, the real space distribution of quantum particles in a topological atom, $\Omega$, is completely determined by following probabilities:

$$P_n^\pm(\Omega) = \binom{N_\pm}{n} \int_\Omega d\vec{r}_1^\pm \cdots \int_\Omega d\vec{r}_n^\pm \int_{R^3-\Omega} d\vec{r}_{n+1}^\pm \cdots \int_{R^3-\Omega} d\vec{r}_{N_\pm}^\pm \; \gamma^{N_\pm}\left(\vec{r}_1^\pm, \cdots, \vec{r}_{N_\pm}^\pm\right) \qquad (4)$$

$$\gamma^{N_\pm}\left(\vec{r}_1^\pm, \cdots, \vec{r}_{N_\pm}^\pm\right) = \sum_{spins} \int d\vec{r}_1^\mp \cdots \int d\vec{r}_{N_\mp}^\mp \; \Psi^*\left(\vec{x}_1'^-, \cdots, \vec{x}_{N_-}'^-, \vec{x}_1'^+, \cdots, \vec{x}_{N_+}'^+\right) \Psi\left(\vec{x}_1^-, \cdots \vec{x}_{N_-}^-, \vec{x}_1^+, \cdots, \vec{x}_{N_+}^+\right) \Big|_{\vec{x}_i' = \vec{x}_i}$$

In these equations $\gamma^{N_\pm}$ is the *analogues* of the diagonal elements of the orthodox spinless $N_\pm$-order density matrices (the summation over the spin variables transforms the original variables, $\vec{x}_i^\pm = \vec{r}_i^\pm . \vec{\sigma}_i$, into vectors in real space, $\vec{r}_i^\pm$) [47], while $n$ stands for the number of a certain type of quantum particles residing in atomic basin, $\Omega$. Each $P_n^\pm(\Omega)$ is the probability of an *event*; the probability of observing a certain configuration that $n$ indistinguishable quantum particles of a certain type are in basin $\Omega$ whereas the others, $N_\pm - n$, are located in remaining basins, $R^3 - \Omega$. These probabilities are the basic ingredients of the real space distribution of quantum particles. It is straightforward to demonstrate that for a normalized wavefunction:

$$\sum_n P_n^\pm(\Omega) = 1$$

$$N_\pm(\Omega) = \sum_n n P_n^\pm(\Omega) \tag{5}$$

In these equations $N_\pm(\Omega)$ is the means (first statistical moments) of the real space distributions termed populations of quantum particles in atomic basins, $N_\pm(\Omega) = \int_\Omega d\vec{r}_\pm \, \rho_\pm(\vec{r}_\pm)$ [12]. The second statistical moment is the variance of the real space distribution:

$$\Lambda_\pm(\Omega) = \overline{N_\pm^2(\Omega)} - [N_\pm(\Omega)]^2 = \sum_n n^2 P_n^\pm(\Omega) - \left[\sum_n n P_n^\pm(\Omega)\right]^2 \tag{6}$$

In statistical mechanics this quantity is a measure of the *fluctuations* in the number of particles of open systems defined for the grand canonical ensemble [48]. The same interpretation is also employed in considering the *real space open quantum subsystems* [5]; $\Lambda_\pm(\Omega)$ of an atomic basin quantify the amount of *quantum fluctuations* of both electrons and the PCPs [37,39]. Consequently, for a *closed* subsystem $\Lambda_\pm(\Omega)$ approaches zero although this is a hypothetical state not fully realized in molecular systems [43].

Taking this background into account, one may claim that the amount of $\Lambda_\pm(\Omega)$ is a natural measure of localization/delocalization; roughly speaking, a small $\Lambda_\pm(\Omega)$ indicates higher localization whereas a large $\Lambda_\pm(\Omega)$ manifests a pronounced delocalization. In order to explicitly determine this quantity, one must compute $\overline{N_\pm^2(\Omega)}$ in equation (6) that needs introducing the extended reduced second order spinless density matrices [10]:

$$\rho_\pm^{(2)}(\vec{q}_1,\vec{q}_2) = N_\pm(N_\pm - 1) \int d\vec{r}_3^\pm \ldots \int d\vec{r}_{N_\pm}^\pm \, \gamma^{N_\pm}\left(\vec{r}_1^\pm,\cdots,\vec{r}_{N_\pm}^\pm\right) \bigg|_{\vec{r}_1^\pm = \vec{q}_1, \vec{r}_2^\pm = \vec{q}_2} \tag{7}$$

Based on this definition, it is straightforward to demonstrate:

$$\Lambda_\pm(\Omega) = \int_\Omega d\vec{q}_1 \int_\Omega d\vec{q}_2 \, \rho_\pm^{(2)}(\vec{q}_1,\vec{q}_2) + N_\pm(\Omega) - [N_\pm(\Omega)]^2 \tag{8}$$

In order to proceed, based on previous proposals [5,47], the extended reduced second order spinless density matrices is decomposed as follows:

$$\rho_{\pm}^{(2)}(\vec{q}_1,\vec{q}_2) = \rho_{\pm}(\vec{q}_1)\rho_{\pm}(\vec{q}_2)\left[1 + f_{\pm}(\vec{q}_1,\vec{q}_2)\right] \tag{9}$$

In this equation $f_{\pm}(\vec{q}_1,\vec{q}_2)$ captures the *correlated* and *self-counting* contributions to the density matrix (many authors in the last four decades have concerned on the nature and the information which is contained in this function and there is a large literature devoted to these developments, however, in this contribution, this function is considered very briefly leaving a thorough analysis to a future contribution) [5,37,39]. According to this decomposition, equation (8) is rewritten as follows:

$$\Lambda_{\pm}(\Omega) = F_{\pm}(\Omega,\Omega) + N_{\pm}(\Omega)$$

$$F_{\pm}(\Omega,\Omega) = \int_{\Omega} d\vec{q}_1 \int_{\Omega} d\vec{q}_2 \ \rho_{\pm}(\vec{q}_1)\rho_{\pm}(\vec{q}_2) f_{\pm}(\vec{q}_1,\vec{q}_2) \tag{10}$$

Generally, $|F_{\pm}(\Omega,\Omega)| \prec N_{\pm}(\Omega)$, thus for an open subsystem the inequality $\Lambda_{\pm}(\Omega) \succ 0$ holds. It is customary to introduce the *localization index (LI)*: $\lambda_{\pm}(\Omega) = |F_{\pm}(\Omega,\Omega)|$ [39]. As final step, one may introduce the *relative fluctuation* and the *percent localization*:

$$\frac{\Lambda_{\pm}(\Omega)}{N_{\pm}(\Omega)} = 1 + \frac{F_{\pm}(\Omega,\Omega)}{N_{\pm}(\Omega)} \quad , \quad Ploc_{\pm}(\Omega) = \left|\frac{F_{\pm}(\Omega,\Omega)}{N_{\pm}(\Omega)}\right| \times 100 \tag{11}$$

Since $0 \prec \dfrac{\Lambda_{\pm}(\Omega)}{N_{\pm}(\Omega)} \prec 1$, the percent localization obeys: $0 \prec Ploc_{\pm}(\Omega) \prec 100$; for electrons both limits are unattainable, however, for massive quantum particles one may observe near perfect localization as is demonstrated in subsequent subsection.

In order to quantify delocalization, the population fluctuations of two atomic basins, $\Omega$ and $\Omega'$, are considered as a joint open quantum subsystem, $\Omega'' = \Omega + \Omega'$. It is straightforward to derive:

$$\Lambda_{\pm}(\Omega'') = \Lambda_{\pm}(\Omega) + \Lambda_{\pm}(\Omega') + F_{\pm}(\Omega,\Omega')$$

$$F_{\pm}(\Omega,\Omega') = \int_{\Omega} d\vec{q}_1 \int_{\Omega'} d\vec{q}_2 \ \rho_{\pm}(\vec{q}_1)\rho_{\pm}(\vec{q}_2) f_{\pm}(\vec{q}_1,\vec{q}_2) \qquad (12)$$

These equations clearly demonstrate that the population fluctuation of the joint basin is *not* the sum of the fluctuations of each atomic basin; roughly speaking, $F_{\pm}(\Omega,\Omega')$ is a measure of *inter-basin* contribution that originates from delocalization of particles between the two basins. One may clearly interpret this inter-basin contribution as a measure of the delocalization of quantum particles residing in basin $\Omega$ into the basin $\Omega'$ and vice versa. This interpretation is also in line with the fact that the total population for each atomic basin, for a system composed of an arbitrary number of basins, is just as follows:

$$N_{\pm}(\Omega) = -\left\{ F_{\pm}(\Omega,\Omega) + \sum_{\Omega'} F_{\pm}(\Omega,\Omega') \right\} \qquad (13)$$

This equation demonstrates that apart from the sign, the population of each basin is the sum of localized and delocalized populations. Similar to the *LI*, it is customary to introduce the *delocalization index (DI)*: $\delta_{\pm}(\Omega,\Omega') = |F_{\pm}(\Omega,\Omega') + F_{\pm}(\Omega',\Omega)|$ [43].

The basic ingredients of the extended localization/delocalization analysis are now at hand and it is possible to compute relevant quantities from ab initio wavefunctions, however, before such considerations, it is instructive to introduce simple models to illustrate the nature of localization/delocalization of the PCPs and their mass dependence in the considered four-body systems.

**3.2 Mass dependence of localization/delocalization indexes**

In order to have a qualitative understanding of the ab initio trends as well as the origin of the PCPs' localization/delocalization, simple models describing the four body systems, composed of two atomic basins (*vide infra*), are introduced in this subsection. Accordingly, at first step, based on the mean-field idea, the total wavefunction is assumed to be the product of electronic wavefunction and a Hartree product (HP) of PCPs; the use of the HP is a reasonable approximation in the case of massive localized particles (see [49] for a detailed discussion as well as citations of relevant previous works). However, one expects that upon the decrease of the mass, this approximation deteriorates and as will be discussed, determinants/permanents must be replaced by the HP for indistinguishable fermions/bosons. Thus, deviations of ab initio computed localization/delocalization indexes based on determinants/permanents from those predicted by the model based on HP are a measure of the role of particles' *exchange* in the case of fermions as well as the importance of the superposition of localized orbitals describing the PCPs (*vide infra*). In this regard, two spatial orbitals for the PCPs are introduced, $\psi_i : i = 1,2$, each composed of a single normalized Gaussian function located in an atomic basin, $\psi_i = \phi_i(\vec{r}_i^+ - \vec{R}_i) = (2\alpha/\pi)^{3/4} Exp\left[-\alpha(\vec{r}_i^+ - \vec{R}_i)^2\right]$, in order to compute $F_+(\Omega,\Omega)$ using proper second order spinless density matrices previously derived for the HP [50]:

$$F_+(\Omega,\Omega) = -S_{11}^2(\Omega) - S_{22}^2(\Omega)$$

$$F_+(\Omega,\Omega') = -S_{11}(\Omega)S_{11}(\Omega') - S_{22}(\Omega)S_{22}(\Omega')$$

$$S_{ii}(\Omega) = \int_\Omega d\vec{r}_i^+ \psi_i^*(\vec{r}_i^+) \psi_i(\vec{r}_i^+) \qquad (14)$$

According to equations (10) and (12) the *LI* and *DI* are derived based on the overlap integrals:

$$\lambda_+(\Omega) = S_{11}^2(\Omega) + S_{22}^2(\Omega)$$

$$\delta_+(\Omega,\Omega') = 2\left|-S_{11}(\Omega)S_{11}(\Omega') - S_{22}(\Omega)S_{22}(\Omega')\right| \qquad (15)$$

In order to proceed, the regional overlap integrals, $S_{ii}(\Omega)$, must be evaluated explicitly. Locating the center of Gaussian functions at $\vec{R}_i = (0,0,(-1)^i d/2)$, with identical exponents, $\alpha$, and assuming $xy$ plane as the inter-atomic surface it is straightforward to demonstrate:

$$S_{11}(\Omega) = S_{22}(\Omega') = (1/2)\{1 + Erf[d\sqrt{\alpha/2}]\}$$

$$S_{22}(\Omega) = S_{11}(\Omega') = (1/2)\{1 - Erf[d\sqrt{\alpha/2}]\} \tag{16}$$

In these equations *Erf* stands for the *error function*. By substituting equations (16) into equations (15) the *LI/DI* are derived:

$$\lambda_+(\Omega) = (1/2)\{1 + Erf^2[d\sqrt{\alpha/2}]\}$$

$$\delta_+(\Omega,\Omega') = 1 - Erf^2[d\sqrt{\alpha/2}] \tag{17}$$

These equations are illustrative since the relevant variables are only the distance between the center of the two Gaussian functions as well as the exponent; based on variational optimization of total energy, it is revealed that both of these quantities, as variational parameters, are *mass dependent*. According to the derived ab initio computed exponents and distances which to be discussed in subsequent sections, Table 1, one may propose following equations to quantify the mass dependence:

$$d(m_+) = am_+^{-n} + b$$

$$\alpha(m_+) = (a' + b'm_+^p)^2 \qquad n, p \succ 0 \tag{18}$$

In these equations all parameters, $a, b, n, a', b', p$, are determined by non-linear regression, employing the data from the mass variation range: $m_+ = 10m_e - 2000m_e$, yielding: $a \approx 3.8192$, $b \approx 1.3835$, $n \approx 0.5035$, $a' \approx -0.7877$, $b' \approx 0.8439$, $p \approx 0.2500$ (Figures S1 and S2 in Supporting Information depict both the used ab initio data as well as equations (18)).

Substituting equations (18) into equations (17), assuming for simplicity $n=1/2$ and $p=1/4$, yields the final mass-dependent *LI/DI*:

$$\lambda_+(\Omega;m_+) = (1/2)\left\{1 + Erf^2\left[g_1 + g_2 m_+^{1/4} + g_3 m_+^{-1/4} + g_4 m_+^{-1/2}\right]\right\}$$

$$\delta_+(\Omega,\Omega';m_+) = 1 - Erf^2\left[g_1 + g_2 m_+^{1/4} + g_3 m_+^{-1/4} + g_4 m_+^{-1/2}\right] \qquad (19)$$

In these equations the set of $\{g_i\}$ are known coefficients ($g_1 = a'b/\sqrt{2} \prec 0$, $g_2 = bb'/\sqrt{2} \succ 0$, $g_3 = ab'/\sqrt{2} \succ 0$, $g_4 = aa'/\sqrt{2} \prec 0$). The explicit mass dependence of *LI/DI* is a novel feature of the TC-QTAIM that has been also demonstrated recently in the case of various basin properties [10]. Upon the increase of the mass, say for $m_+ \geq 400 m_e$, for all practical proposes only the second term in the argument of error function is of importance and one may further simplify the equations:

$$\lambda_+(\Omega;m_+) \approx (1/2)\left\{1 + Erf^2\left[g_1 + g_2 m_+^{1/4}\right]\right\}$$

$$\delta_+(\Omega,\Omega';m_+) \approx 1 - Erf^2\left[g_1 + g_2 m_+^{1/4}\right] \qquad (20)$$

These equations, which may be used directly for nuclei, yield the *proper* limits for massive particles:

$$\lim_{m_+ \to \infty} \lambda_+(\Omega;m_+) \to 1$$

$$\lim_{m_+ \to \infty} \delta_+(\Omega,\Omega';m_+) \to 0 \qquad (21)$$

Therefore, as one expects, the model based on the HP predicts that each massive PCP is totally localized in a single atomic basin with no delocalization into neighboring basin. Figure 1a depicts equations (19) as the *LI/DI* deduced from the model versus the mass of the PCPs; it is evident from this figure that only for $m_+ \prec 50 m_e$ the delocalization of the PCPs is of relevance. It is timely to emphasize that the origin of the delocalization is the *leakage* of the quantum wave of a PCP, located primarily in its basin, into neighboring basin. Therefore, the *quantum exchange* of identical particles and associated Fermi-hole have nothing to do with this kind of delocalization [37,39].

In order to take explicitly the indistinguishability of the PCPs into account, one needs to construct a $2\times 2$ determinant/permanent constructed from the spin-orbitals of the PCPs. However, since it was demonstrated that the indexes deviate from their limiting value in equation (21) for masses much smaller than those of typical nuclei, just determinants, as fermions' representatives, are considered in this contribution taking light leptons like positrons and muons into account leaving the consideration of permanents/bosons to a future study. Since spin states are of relevance for fermions, both the singlet/para and the triplet/ortho states/species are scrutinized for the considered two-particle case. Let's first consider the para species.

Accordingly, to construct spin-orbitals with opposite spin states, in contrast to the typical two-electron case, two *different* spatial orbitals for the PCPs are introduced, $\psi_i : i = 1,2$, composed of linear combinations of two normalized Gaussian functions each of which located in a single basin, $\psi_i(\vec{r}_i^{\,+}) = c_1^i \phi_1(\vec{r}_i^{\,+} - \vec{R}_i/2) + c_2^i \phi_2(\vec{r}_i^{\,+} + \vec{R}_i/2)$. It is straightforward to demonstrate that equations (14) and (15) hold also for para species while from the ab initio calculations, Table 3, one observes: $c_1^1 = c_2^2$, $c_2^1 = c_1^2$. After some mathematical manipulations, similar to those described for the HP case, and using the fact that $S_{11}(\Omega) = S_{22}(\Omega')$ and $S_{22}(\Omega) = S_{11}(\Omega')$, following equations emerge:

$$\lambda_+(\Omega) = (1/2)\left\{1 + \frac{Erf^{\,2}\left[d\sqrt{\alpha/2}\right]}{1 - Exp\left[-\alpha d^{\,2}\right]}\right\}$$

$$\delta_+(\Omega,\Omega') = 1 - \frac{Erf^{\,2}\left[d\sqrt{\alpha/2}\right]}{1 - Exp\left[-\alpha d^{\,2}\right]} \tag{22}$$

In comparison with equations (17), just the denominators of the second term in the right-hand side of these equations differ from the HP case. However, for large mass region, based on equations (18), it is easy to deduce: $\alpha d^2 \propto \sqrt{m_+}$, and one may conclude that for all practical purposes the denominators disappear and equations (22) reduce to their HP counterparts. This also conforms well to the ab initio derived orbital coefficients (Table 3) since for large mass regime one observes: $c_1^1 = c_2^2 \approx 1$, $c_2^1 = c_1^2 \approx 0$. In other words, in line with the HP wavefunction, each spatial orbital in this region, starting at $m_+ \succ 600 m_e$, is composed of a single Gaussian function located in a single basin. Accordingly, for massive particles, equations (22) yield the *proper* limits, satisfying equations (21), demonstrating the complete localization of each massive particle in a single basin. In order to compare equations (22) with equations (17), similar to the HP case, non-linear regressions were done to deduce the mass-dependence of the ab initio exponents and distances of Gaussian functions, $\alpha(m_+)$ and $d(m_+)$, using equations (18). Employing the ab initio data in the mass region: $m_+ = 10 m_e - 2000 m_e$, Table 3, all parameters, $a, b, n, a', b', p$, are determined using non-linear regressions, yielding: $a \approx 3.4893$, $b \approx 1.3721$, $n \approx 0.4769$, $a' \approx -0.7967$, $b' \approx 0.8487$, $p \approx 0.2495$ (Figures S3 and S4 in Supporting Information depict both the used ab initio data as well as equations (18)). Substituting equations (18) with the derived parameters into equations (22), the explicit mass dependence of the *LI/DI* is revealed as presented graphically in Figure 1b; in line with the HP case, only for $m_+ \prec 50 m_e$ the delocalization of the PCPs is appreciable. Let's now consider the ortho species.

In contrast to the para species, in the ortho species the spin states are the same thus the orthogonality of spin-orbitals dictates the orthogonality of the PCPs' spatial orbitals:

$$S_{12}(R^3) = \int d\vec{r}\, \psi_1^*(\vec{r})\psi_2(\vec{r}) = 0.$$
However, for each basin the regional overlap integral, $S_{12}(\Omega) = \int_\Omega d\vec{r}_i^+ \psi_1^*(\vec{r}_i^+) \psi_2(\vec{r}_i^+)$, is not null and one finds:

$$F_+(\Omega,\Omega) = -S_{11}^2(\Omega) - S_{22}^2(\Omega) - 2S_{12}^2(\Omega)$$

$$F_+(\Omega,\Omega') = -S_{11}(\Omega)S_{11}(\Omega') - S_{22}(\Omega)S_{22}(\Omega') - 2S_{12}(\Omega)S_{12}(\Omega') \qquad (23)$$

In comparison with equations (14), the third term in the right-hand side of these equations distinguishes them from the equations associated with the singlet and the HP states. According to equations (10) and (12), the *LI /DI* are derived as follows:

$$\lambda_+(\Omega) = S_{11}^2(\Omega) + S_{22}^2(\Omega) + 2S_{12}^2(\Omega)$$

$$\delta_+(\Omega,\Omega') = 2\left| -S_{11}(\Omega)S_{11}(\Omega') - S_{22}(\Omega)S_{22}(\Omega') - 2S_{12}(\Omega)S_{12}(\Omega') \right| \qquad (24)$$

One may further simplify these equations using following relations: $S_{11}(\Omega) = S_{22}(\Omega')$, $S_{22}(\Omega) = S_{11}(\Omega')$ and $S_{12}(\Omega') = -S_{12}(\Omega)$; however, before proceeding further, the explicit form of spatial orbitals is desired. Accordingly, employing the same form of spatial orbitals used to describe para species, $\psi_i(\vec{r}_i^+) = c_1^i \phi_1(\vec{r}_i^+ - \vec{R}_i/2) + c_2^i \phi_2(\vec{r}_i^+ + \vec{R}_i/2)$, the ab initio calculations, Table 3, reveal two alternatives for the coefficients of the used Gaussian functions: $c_1^1 = c_2^1$, $c_2^2 = -c_1^2$ for $m_+ \leq 200 m_e$ and $c_1^1 = -c_2^2$, $c_2^1 = c_1^2$ for $m_+ \geq 600 m_e$ (the apparent "mass gap" emerges since the corresponding ab initio calculations were done in discrete mass steps. See the section on ab initio calculations for details). After some mathematical manipulations, similar to those discussed previously, one arrives:

$m_+ \leq 200 m_e$:

$$\lambda_+(\Omega) = (1/2)\left\{1 + \frac{Erf^2\left[d\sqrt{\alpha/2}\right]}{1 - Exp[-\alpha d^2]}\right\}$$

$$\delta_+(\Omega, \Omega') = 1 - \frac{Erf^2\left[d\sqrt{\alpha/2}\right]}{1 - Exp[-\alpha d^2]}$$

$m_+ \geq 600 m_e$:

$$\lambda_+(\Omega) = (1/2)\left\{1 + Erf^2\left[d\sqrt{\alpha/2}\right] + Exp[-\alpha d^2]\right\}$$

$$\delta_+(\Omega, \Omega') = 1 - Erf^2\left[d\sqrt{\alpha/2}\right] - Exp[-\alpha d^2] \tag{25}$$

While the equations for $m_+ \leq 200 m_e$ are the same as equations (22) derived for ortho species, for $m_+ \geq 600 m_e$ the equations are novel though, $\alpha d^2 \propto \sqrt{m_+}$ as discussed previously, and the exponential term in the right-hand side of these equations disappears in the high mass regime and these equations also reduce to their HP counterparts namely, equations (17). Consequently, in the large mass region once again equations (21) hold and the nuclei are completely localized in their associated basins. It is timely to emphasize that this observation as well as the similar fact emerging from considering the large mass regime of para species all demonstrate that the HP model is a legitimate approximation when considering nuclei without any significant intra-molecular tunneling. On the other extreme, in order to scrutinize the difference in the *LI/DI* indexes with the para species and the HP at low mass region, employing equations (18), the parameters of these equations were derived using non-linear regression to the ab initio computed exponents and distances of Gaussian functions, Table 3. The resulting parameters are: $a \approx 2.7959$, $b \approx 1.3383$, $n \approx 0.4107$, $a' \approx -0.8240$, $b' \approx 0.8628$, $p \approx 0.2479$ (Figures S5 and S6 in Supporting Information depict both the used ab initio data as well as equations (18)); substituting equations (18) with the derived parameters into equations (25) for $m_+ \leq 200 m_e$, the explicit mass dependence of *LI/DI* is exposed as offered graphically in Figure 1c. Evidently, it is observed that only for $m_+ \prec 50m_e$ the delocalization of the PCPs is appreciable. As final analysis, it is instructive to compare the three calculated curves of the *LI/DI* indexes versus the PCPs' masses for the considered cases more closely as depicted in Figures 1d and 1e. It is evident from these figures that for the most of the considered masses the differences are quite negligible and the simpler equations of the HP model, equations (19), work quite well. A more thorough comparative analysis on delicate differences is done in subsequent sections on computational studies.

## 4 Computational procedures

The ab initio computations on the four-body systems have been performed by the fully variational multi-component molecular orbital method (FV-MC_MO) developed originally by Tachikawa and coworkers [51,52], employing Gaussian basis sets for both the electrons and the PCPs, [6s:1s] and [5s:1s1p1d] (see [8,53] for a discussion on the employed nomenclature). Twenty masses were selected to scan the entire mass spectrum, $m = 10m_e - 10^{11}m_e$, of the PCPs in discrete steps that are given in Table 1 while the used masses for hydrogen isotopes throughout calculations are: $H = 1836.15267247 m_e$, $D = 3670.4829654 m_e$, $T = 5496.9215269 m_e$. The part of wave function describing the PCPs is treated within various discussed schemes, the HP as well as the usual triplet and singlet states (*vide supra*), while the electronic wavefunction is assumed to be closed shell singlet throughout the whole calculations. In the FV-MC_MO methodology, based on the variational principle, all the variables of the Gaussian functions, i.e. exponents, coefficients and the location of functions, are fully optimized through a non-linear optimization procedure yielding the lowest possible total energy. Accordingly, because of the full variational optimization of the variables of the basis functions, the virial theorem is automatically satisfied thus the computed virial ratios of all species are close to the exact value, $\left(\left\langle \hat{V} \right\rangle / \left\langle \hat{T} \right\rangle\right) = -2 \pm (1 \times 10^{-5})$, relegating the need for the usual *ad hoc* virial scaling of basin energies in the QTAIM analysis [13]. The details of the FV-MC_MO calculations, e.g. one and two-particle integration schemes, problems relevant to numerical stability, non-linear optimization algorithms, SCF convergence accelerators all have been fully disclosed elsewhere and are not reiterated here [8,9]. The corresponding FV-MC_MO code is under constant development dealing with larger many-body systems with a reasonable computational cost, however, one must note that for the latter systems the non-linear optimization is a computationally cumbersome procedure and simplifications are necessary as will be discussed in future contributions. As a reference data set, some calculations done by Tachikawa and Osamura on hydrogen molecule and its isotopomers were redone and successfully reproduced [54]. It is important to realize that in the FV-MC_MO methodology a molecular-fixed frame is always assumed for the system under study, by clamping some nuclei or employing an externally imposed fixed frame, thus avoiding the delicate problem of translational-rotational invariance; this pivotal cornerstone and its consequences on the TC-QTAIM analysis have been fully discussed and elaborated in the previous study [9]. Thus, the resulting wavefunctions used for the subsequent TC-QTAIM analysis correspond to the WF1 family [9]. In this study the z axis is defined as a line going through the center of Gaussian functions describing the PCPs while the middle of this center is used as the center of coordinate system. The associated TC-QTAIM analysis was done by the code described fully in previous communications [8,13]. The added new ability is the calculation of the *LI* and *DI* indexes based on the recipes developed in previous section. In all basin integrations great care was devoted to set the integration parameters properly, thus avoiding incomplete space sampling that is particularly problematic for species containing the lighter PCPs [13].

# 5 The ab initio calculations

The ab initio FV-MC_MO calculations were performed based on the protocol described in previous section and some results are gathered in Tables 1, 2 and 3. Inspection of Table 1 reveals that all computed properties, energy and its components as well as mean inter-particle distances, using the HP wavefunction and [6s:1s] basis set, smoothly vary upon increasing the mass of the PCPs and in the large mass limit converge toward corresponding values computed independently within the clamped nucleus model (see the last row of the table). Accordingly, based on equations (18) the exponent of the Gaussian functions describing the PCPs increases with the mass, $\alpha \propto \sqrt{m_+}$ for large masses, thus for larger masses the Gaussian functions tend to Dirac delta functions. The latter is a mathematical manifestation of physically localized particles [9,10] and in line with this observation the kinetic energy of the PCPs tends to zero for large masses thus for all practical purposes masses with $m \geq 10^{11} m_e$ are just clamped particles. This mass-induced localization manifests itself also in other energy components; roughly speaking, because of the gradual localization of the PCPs as well as contraction of the inter-particle distance, electrons circulate around the PCPs in orbits with smaller radius occupying a more compact region thus electronic kinetic energy as well as the absolute magnitude of electrons-PCPs stabilizing potential energy increases upon the increase of the mass. On the other hand, the contraction of electrons' as well as the PCPs' distributions elevates both the destabilizing electron-electron and PCP-PCP potential energy interactions. Table 2 offers the results of ab initio calculations using the HP wavefunction and employing the more extended basis set, [5s:1s1p1d]. The general regularities observed in this table are the same as those of Table 1 and are not reiterated, however, it is noticeable that the total energy associated with each mass is lower than the corresponding total energy in Table 1 although, as presented in the last column of this table, the energy difference of these two diminishes upon increasing the mass. The latter is explicable if one notes that anharmonic vibrational dynamic, described by *p* and *d*-type Gaussian functions, is less pronounced for larger masses. Accordingly, beyond $m \geq 10^5 m_e$ (data not included in the table), the total energies computed by [6s:1s] basis set are lower than those computed by [5s:1s1p1d] basis set demonstrating the dominance of the role of the single extra electronic basis function, [6s] versus [5s], in the variational optimization of the total energy. Table 3 that contains the results of ab initio calculations, using the wavefunction composed of Slater determinants with [6s:1s] basis set, also discloses the same general regularities and patterns observed in Table 1 for both the singlet and triplet states thus they are not reiterated. On the other hand, comparative analysis of Tables 1 and 3 casts no doubt that even with a conservative evaluation, for $m \geq 100 m_e$, the results of the ab initio calculations are the same for the three employed wavefunctions further demonstrating the utility of the HP wavefunction, as a simplified wavefunction, detailed elsewhere [49]. Accordingly, the computed overlap integral of the two Gaussian functions in the whole space diminishes rapidly upon increasing the mass and except for $m = 10 m_e, 50 m_e$ (with overlap integrals ~0.17 and ~0.02, respectively), it is completely negligible; with no such overlap, the PCPs are effectively distinguishable particles and the use of the HP wavefunction is justified.

In subsequent section, the derived wavefunctions are used for the TC-QTAIM analysis though based on what emerged from the ab initio calculations one expects, neglecting $m \prec 100 m_e$, that the main features of the analysis to be the same for the three employed wavefunctions as far as the same basis set is used; particularly, this promises that even for the lightest nuclei namely hydrogen isotopes, the analysis being insensitive to the used wavefunctions.

## 6 The TC-QTAIM analysis

Similar to its orthodox analogue [5-7], the TC-QTAIM analysis is composed of two cornerstones namely, the topological analysis of $\Gamma^{(2)}(\vec{q})$ deciphering the morphology of AIM and computing the regional properties of AIM. In forthcoming two subsections, these ingredients of the TC-QTAIM are implemented on the considered four-body systems while the last subsection contains the results of numerical calculations on the *LI/DI*. Also, in discussing the patterns of various TC-QTAIM derived quantities in forthcoming subsections, it is implicitly assumed that all regularities and trends are described from low to high masses thus the phrase "due to the increase of the mass" is eliminated from corresponding statements.

### 6.1 Topological analysis

The topological analysis of $\Gamma^{(2)}(\vec{q})$ on all considered species, regardless of the used wavefunction and basis set, reveals a familiar topological structure namely, two global attractors, (3, -3) critical points (CPs), with the same topological properties and a single bond critical point (BCP), (3, -1) CP, between them as is evident from Figures 2 and 5. This observation demonstrates that each species, regardless of the mass of the PCPs, is composed of two equivalent AIM (like the usual homo-nuclear diatomics). Some results of the performed topological analysis at the (3, -3) and (3, -1) CPs, using the HP wavefunction and [6s:1s] basis set, are gathered in Table 4 while Figures 2-5 depict the relief maps of $\Gamma^{(2)}(\vec{q})$ as well as its "components" namely, $\rho_+(\vec{q})$ and $\rho_-(\vec{q})$, for selected masses. It is evident from Table 4 and Figure 2 that the distance between (3, -3) and (3, -1) CPs (usually called bond path within jargon of the orthodox QTAIM) decreases disclosing the "contraction" of $\Gamma^{(2)}(\vec{q})$; in line with this contraction, the absolute values of $\Gamma^{(2)}(\vec{q})$ at CPs move up in this series of species. A detailed inspection however demonstrates that the concentration of $\Gamma^{(2)}(\vec{q})$ around (3, -3) CPs are more pronounced and this trend is best illustrated considering: $\Gamma^{(2)}((3,-1)CP)/\Gamma^{(2)}((3,-3)CP)$; this ratio decreases smoothly from ~0.92 for $m=10m_e$ to ~0.64 for $m=10^7 m_e$. It is tempting to interpret this trend as a sign of a *continuous transition from more floppy to more rigid topological structures*; increasing the mass induces a more pronounced concentration of electrons as well as the PCPs at the global attractors amplifying the "clustering" of quantum particles around these points [55]. Evidently, the observed AIM structure is independent from the BO approximation as well as the mass ratio.

At this stage it is illustrative to consider the independent topological analysis of the components of $\Gamma^{(2)}(\vec{q})$ thus Tables 5 and 6 as well as Figures 2-5 contain some relevant data and relief maps. Inspection of Table 5 and Figure 3 reveals that $\rho_+(\vec{q})$ has the same topological structure of $\Gamma^{(2)}(\vec{q})$ as well as a strong "contraction" trend; this is true regardless of the used wavefunction or basis set. As a general trend, the concentration of $\rho_+(\vec{q})$ around its (3, -3) CPs grows rapidly disclosing the similar trend previously mentioned for the localization of the PCPs' distribution. For $m \geq 600m_e$ the value of $\rho_+(\vec{q})$ at its (3, -1) CP is practically null; at global attractors, the absolute values of $\nabla^2 \rho_+(\vec{q})$, as a proper measure of local concentration [5,6], grows rapidly resembling the trait of Dirac delta function while at BCPs $\Gamma^{(2)}(\vec{q}) = \rho_-(\vec{q})$. Inspection of Table 6 and Figure 4 reveals that for $m \geq 50m_e$ the topological structure of $\rho_-(\vec{q})$ is the same as that derived for $\Gamma^{(2)}(\vec{q})$ and $\rho_+(\vec{q})$ even so for $m=10m_e$ just a single global attractor is observed at the center of coordinate system; as is evident from Figure 5, for $m=10m_e$, what shapes the topological structure of $\Gamma^{(2)}(\vec{q})$ is $\rho_+(\vec{q})$ *not* $\rho_-(\vec{q})$. Although, in contrast to the case of $\Gamma^{(2)}(\vec{q})$ and $\rho_+(\vec{q})$, the length of bond paths increases throughout the series, the observed trend of $\rho_-(\vec{q})$ and $\nabla^2 \rho_-(\vec{q})$ at both the (3, -3) and (3, -1) CPs signifies the general contraction of $\rho_-(\vec{q})$; the ratio $\rho_-((3, -1)CP)/\rho_-((3, -3)CP)$ decreases smoothly from ~0.97 for $m = 50m_e$ to ~0.64 for $m = 10^7 m_e$. This is comprehensible since a locally concentrated/localized one-density of the PCPs, induces a concomitant concentration into the one-density of electrons. A more detailed inspection reveals that the position of each (3, -3) CP of $\Gamma^{(2)}(\vec{q})$ for $m \leq 10^7 m_e$ is in a good approximation the arithmetic mean of position of (3, -3) CPs of $\rho_+(\vec{q})$ and $\rho_-(\vec{q})$ while for $m \geq 10^8 m_e$ the position of (3, -3) CPs of $\Gamma^{(2)}(\vec{q})$ is virtually the same with the position of (3, -3) CPs of $\rho_-(\vec{q})$. Evidently, for $m \leq 10^7 m_e$, in contrast to the properties of the BCP of $\Gamma^{(2)}(\vec{q})$ and points around, $\rho_+(\vec{q})$ actively participates in shaping the morphology of $\Gamma^{(2)}(\vec{q})$ around (3, -3) CPs while for $m \geq 10^8 m_e$ one observes: $\Gamma^{(2)}(\vec{q}) = \rho_-(\vec{q})$. This is rationalized according to equation (1) and Figure 5, since the role of the mass scaled positive density, $(m_-/m_+)\rho_+(\vec{q})$, contributing in shaping $\Gamma^{(2)}(\vec{q})$ is diminishing throughout series; in the infinite mass limit $\rho_-(\vec{q})$ is indistinguishable from the usual one-electron density, $\rho_e(\vec{q})$, used for the topological analysis within context of the orthodox QTAIM thus: $\lim_{m \to \infty} \Gamma^{(2)}(\vec{q}) \to \rho_e(\vec{q})$ [9,10].

The topological analysis not only consists of considering $\Gamma^{(2)}(\vec{q})$ and its derivatives but also derives local property densities, introduced by equation (2), at CPs usually called topological indexes [8]. Accordingly, Table 4 offers some local energy densities as typical property densities that include the Hamiltonian, $\tilde{K}(\vec{q}) = K_+(\vec{q}) + K_-(\vec{q})$, and the Lagrangian, $\tilde{G}(\vec{q}) = G_+(\vec{q}) + G_-(\vec{q})$, kinetic energy densities as well as the total virial field, $\tilde{V}^T(\vec{q}) = V_+^T(\vec{q}) + V_-^T(\vec{q})$, all introduced in detail previously [10]. The trends observed for the absolute values of these property densities as well as $\nabla^2 \Gamma^{(2)}(\vec{q})$ at the global attractors are similar, including an ascending region until $m = 10^7 m_e$ with a sudden drop at $m = 10^8 m_e$, and then a smooth convergence toward the corresponding values computed within the context of the orthodox QTAIM. The latter observation, also demonstrated in a previous analytical study [10], is in line with the general trait of property densities introduced within the TC-QTAIM: $\lim_{m\to\infty} \tilde{M}(\vec{q}) \to M(\vec{q})$, where $M(\vec{q})$ is the property density computed independently employing the orthodox QTAIM. The observed trends of the topological indexes at the (3, -1) CPs are more or less (but not quite) smooth, converging toward the QTAIM values computed within the framework of the clamped nucleus model. In order to rationalize these trends, topological analysis of $\rho_+(\vec{q})$ and $\rho_-(\vec{q})$ were done and corresponding topological indexes were also computed. Inspection of Table 5 unravels the fact that at BCPs, for $m \geq 600 m_e$, $M_+(\vec{q})$ is null thus just electrons are contributing to the topological indexes while comparison of Tables 4 and 6 demonstrates that at BCPs: $\tilde{M}(\vec{q}) = M_-(\vec{q})$, thus the relatively smooth convergence of $M_-(\vec{q})$ at BCPs throughout the series guarantees the same pattern for $\tilde{M}(\vec{q})$. On the other hand, Table 5 makes known that the trends of the absolute values of $\rho_+(\vec{q})$, $\nabla^2 \rho_+(\vec{q})$, $K_+(\vec{q})$ and $V_+^T(\vec{q})$ at the (3, -3) CPs of $\rho_+(\vec{q})$ all are ascending throughout the series; in a previous study [10], these trends were considered analytically and it was demonstrated that all these quantities ultimately diverge at the (3, -3) CPs of $\rho_+(\vec{q})$ while becoming null in all other points of space. Accordingly, for all points of space except the global attractors of $\rho_+(\vec{q})$ the mass dependence of mentioned densities is as follows: $\sim m^n Exp\left[-a\sqrt{m}\right]$ ($n, a > 0$), thus, based on the explicit form of mass-dependence, one expects first an ascending and then a descending trend throughout the series. Since according to Table 6 the topological indexes at the (3, -3) CPs are smoothly converging toward the values computed within the QTAIM, the non-smooth convergence patterns of the topological indexes at the CPs of $\Gamma^{(2)}(\vec{q})$ must originate from the PCPs' contributions. In order to disentangle the contributions emerging from the PCPs, $M_+(\vec{q})$, and electrons, $M_-(\vec{q})$, to the topological indexes, $\tilde{M}(\vec{q})$, at the (3, -3) CPs of $\Gamma^{(2)}(\vec{q})$, Table S1 in Supporting Information provides each contribution at these points for all the considered species (it is important to realize that since (3, -3) CPs of $\Gamma^{(2)}(\vec{q})$, $\rho_+(\vec{q})$ and $\rho_-(\vec{q})$ are located at different points of z-axis, the data in Tables 5 and 6 can not be used directly for this propose). It is evident from this table that the previously mentioned ascending trend for $\tilde{M}(\vec{q})$ at $m \leq 10^7 m_e$ is dominated/dictated by the ascending trend of $M_+(\vec{q})$ while for $m \geq 10^{10} m_e$, $M_+(\vec{q})$ is practically null and $\tilde{M}(\vec{q}) = M_-(\vec{q})$. This observation clearly lays bare that for a large mass region, including particularly the masses of muon, proton, deuteron and tritium, contributions originating from the PCPs to the topological indexes at (3, -3) CPs, in contrast to BCPs, are even more important compared with those emerging from electrons.

In order to compare with Table 4, Table 7 includes the results of the topological analysis of $\Gamma^{(2)}(\vec{q})$ for singlet/para and triplet/ortho states/species as well as the HP wavefunction with [5s:1s1p1d] basis set for $m \leq 600 m_e$. It is evident from this table that the difference of computed data for singlet and triplet states with corresponding values in Table 4 is slight for $m = 10 m_e, 50 m_e$ and virtually null for $m \geq 100 m_e$ whereas the variations of topological data with the change of basis set, [6s:1s] to [5s:1s1p1d], are more pronounced (though, even in this case, the difference is no more than some percent). All in all, the observed trends in Table 7 are similar to those described in the case of the HP wavefunction and [6s:1s] basis set and are not reiterated; the role of anharmonic nuclear distribution, described by $p$ and $d$-type functions, as well as replacing determinant with HP wavefunction, have seemingly a marginal role on the general aspects of the topological analysis.

**6.2 Basin properties**

Table 8 presents some results of basin integrations, which yield regional properties of AIM, in the considered series of species using the HP wavefunction and [6s:1s] basis set (since all the considered species are like homo-nuclear diatomics, the properties of the both AIM are the same thus a single entry for each basin property is given in the table). Apart from the basin energy and its ingredients as well as polarization dipole, all have been scrutinized in detail in a previous study [10], atomic/basin volumes [56], $V(\Omega) = \int_\Omega k \, d\vec{q}, k = \begin{cases} 1, & if \ \Gamma^{(2)}(\vec{q}) \geq 0.001 \\ 0, & if \ \Gamma^{(2)}(\vec{q}) \prec 0.001 \end{cases}$, and the effective size of the PCP distribution (ESD) [8,9], which is the radius of a sphere located at the (3, -3) CP of $\rho_+(\vec{q})$ containing 0.99999 of PCP's population [8,9], are also included in this table. Inspection of Table 8 demonstrates that basin energies, $\tilde{E}(\Omega) = E_-(\Omega) + E_+(\Omega)$, as well as the components, $E_-(\Omega)$ and $E_+(\Omega)$, converge toward the proper limits [10]: $\lim_{m \to \infty} E_-(\Omega) \to E_e(\Omega)$, $\lim_{m \to \infty} E_+(\Omega) \to 0$, $\lim_{m \to \infty} \tilde{E}(\Omega) = \lim_{m \to \infty} E_-(\Omega) \to E_e(\Omega)$, where $E_e(\Omega)$ is the basin energy computed employing the clamped nucleus model within context of the orthodox QTAIM. Considering the ratio of $E_+(\Omega)/E_-(\Omega)$ in this series reveals an accelerated diminishing trend from ~0.414 for $m = 10 m_e$ to ~$10^{-5}$ for $m = 10^{11} m_e$ while for the hydrogen isotope masses, $m = H, D, T$, this ratio is ~0.036, ~0.026 and ~0.022, respectively, unraveling that the dominate contribution is emerging from electrons. Evidently, the total electric dipole moment of each of the considered species is null; since the both basins contain equal particle populations, $N_\pm(\Omega) = 1$, the charge transfer dipoles are also null and just the polarizations dipoles are considered. Accordingly, assuming the (3, -3) CP of $\rho_+(\vec{q})$ as the center of coordinate system for computing the polarization dipoles, $\vec{P}_\pm(\Omega) = \pm \int_\Omega d\vec{r}^\pm \ \vec{r}^\pm \rho_\pm(\vec{r}^\pm)$, it is straightforward to demonstrate that for the used basis set $\vec{P}_+(\Omega) = 0$ [10], thus just the electronic polarization dipole is considered herein. The electronic polarization dipoles converge smoothly toward the proper limit, computed within the orthodox QTAIM; evidently, the localization of the PCPs throughout the series enhances the electronic polarization. The atomic volumes, in line with the contraction of $\Gamma^{(2)}(\vec{q})$ described in previous subsection, are smoothly descending reaching the proper limit computed independently within context of the orthodox QTAIM. On the other hand, the trend observed for the ESDs of the series also conforms to this picture unraveling the shrinkage of the PCPs' distribution. Interestingly, for $m \leq 600 m_e$ the ESDs are larger than the distance of (3, -3) CPs of $\rho_+(\vec{q})$ from corresponding (3, -1) CPs (see Table 5) thus the ESDs are "overlapping" and the PCP one-densities of each basin not only "reaches" to the (3, -1) CPs but also penetrates into the neighboring basin; in line with this observation, comparison with Table 5 reveals that in the exactly same mass region $\rho_+(\vec{q})$ is non-zero at the (3, -1) CPs. Table 9 offers some results of basin integrations employing the singlet and triplet as well as the HP-[5s:1s1p1d] wavefunctions and as is evident from this table, the general trends are not different from those observed for Table 8 thus they are not reiterated. In the case of the HP-[5s:1s1p1d] wavefunction the polarization dipole of the PCPs is not null revealing a descending trend that goes to zero (as its proper limit) while the PCPs polarization seemingly enhances the concomitant electronic polarization. A comparison with Table 8 once again confirms that just for $m \leq 50 m_e$ the TC-QTAIM analysis of the both singlet and triplet states are different, though slightly, from that of the HP wavefunction.

As the final stage of the analysis, the *LI/DI* are considered for both the electrons and PCPs; the computed regional overlap integrals of electrons in their singlet state, irrespective to the used wavefunction for PCPs, are always 0.5 thus: $\lambda_-(\Omega) = 0.5$ and $\delta_+(\Omega, \Omega') = 1$. On the other hand, Table 10 presents the *LI/DI* for the PCPs employing the four used wavefunctions using equations (15) and (24), directly. In line with Figure 1, descending trends are observed for all the *DI* of the considered cases and for $m \geq 600 m_e$, even with a conservative evaluation, for all practical purposes the PCPs are completely localized in corresponding atomic basins and only for $m \leq 50 m_e$ a slight rise of the *DI* is observed. On the other hand, inspection of the table demonstrates that the variation of the *LI/DI* with wavefunction/basis set is slight and even for $m = 10 m_e$ is not distinct. However, it is timely to emphasize that since the explicit PCP-electron correlation is not taken into account in the wavefunctions of this study [57-60], it is quite probable that for the PCPs, in lower mass region, the *LI/DI* are overestimated/underestimated. Accordingly, advanced future computational studies, producing correlated wavefunctions, must reveal the exact/quantitative mass dependence pattern of the *LI/DI* in these and similar species.

## 7 Prospects

The extended localization/delocalization scheme introduced in this contribution adds a new ingredient to the toolbox of the TC-QTAIM supplementing the previously introduced components [8-11]. While present study demonstrates that the proposed unified methodology yields reasonable results, localization/delocalization of both massive and light particles in/between atomic basins, the truly novel aspects of this methodology reveal itself in systems containing delocalized nuclei; systems with appreciable intra-molecular quantum tunneling of nuclei. This is a promising domain since proton tunneling is a dominate feature of certain seemingly simple but chemically important species that the protonated methane is an illustrative example [61-69]; in such systems, the traditional localization of nuclei completely disappears and the nature of their floppy structure has been a matter of speculations [66]. On the other hand, to have quantitatively accurate non-BO wavefunctions for the computational TC-QTAIM studies, including the explicit correlation of electron-proton/muon/positron from outset is a desirable feature that has been stressed also in recent developments of non-BO ab initio methodologies [57-60,70-77]. The muonic, both positive and negative, species are also interesting targets for computational TC-QTAIM studies since as recently claimed, based on a kinetic study [78], positively charged muon is conceivable as a lighter isotope of hydrogen while the negatively charged muon, based on recent computational studies [79,80], is able to screen one unit of nuclear charge inducing an artificial "transmutation" of elements. All these species as well as other promising targets are under consideration in our lab and the results will be discussed in future contributions.

Acknowledgments

The authors are grateful to the Research Council of Shahid Beheshti Univestity (SBU) for their financial support. The authors are grateful to Masume Gharabaghi and Shahin Sowlati for their detailed reading of a previous draft of paper and helpful suggestions.

**Figure Legends**

**Fig. 1** The mass dependence of the localization and delocalization indexes of the PCPs for $m_+ \leq 200 m_e$. a) For the Hartree product case, equations (19). b) For the para species/singlet states, equations (22). c) For the ortho species/triplet states, equations (25). d) The simultaneous depiction of the localization index for the all three considered wavefunctions. e) The simultaneous depiction of the delocalization index for the all three considered wavefunctions.

**Fig. 2** The relief and counter maps of the Gamma depicted for selected masses employing the HP-[6s:1s] wavefunction.

**Fig. 3** The relief and counter maps of the one-density of the PCPs depicted for selected masses employing the HP-[6s:1s] wavefunction.

**Fig. 4** The relief and counter maps of the one-density of electrons depicted for selected masses employing the HP-[6s:1s] wavefunction.

**Fig. 5** The simultaneous two-dimensional depiction of the Gamma (black curve), one-density of electrons (red curve) as well as the mass scaled (see equation 1) one-density of the PCPs (blue curve) for selected masses on the z-axis that contains all the three critical points, employing the HP-[6s:1s] wavefunction (see text for details).

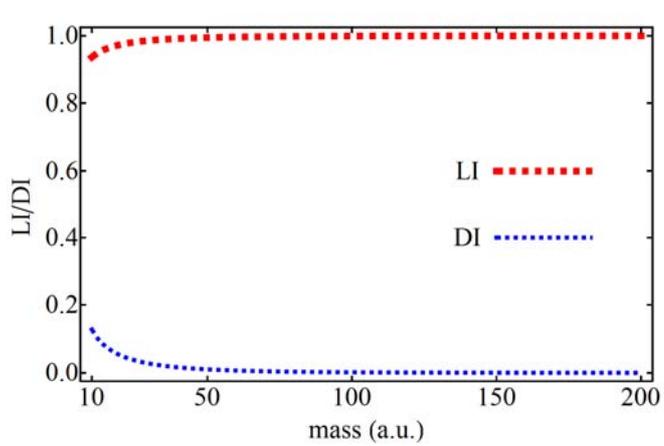

(a)

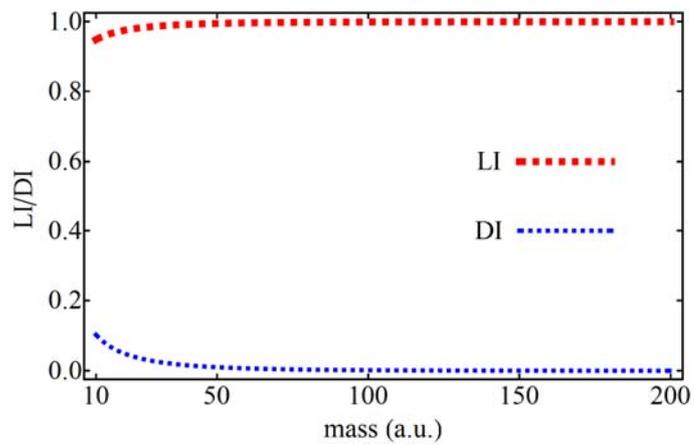

(b)

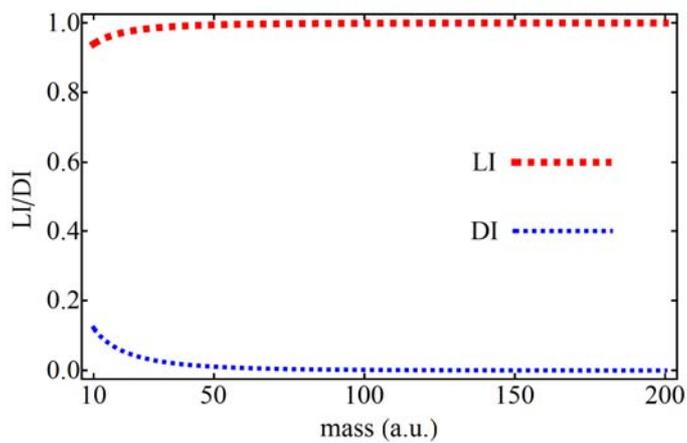

(c)

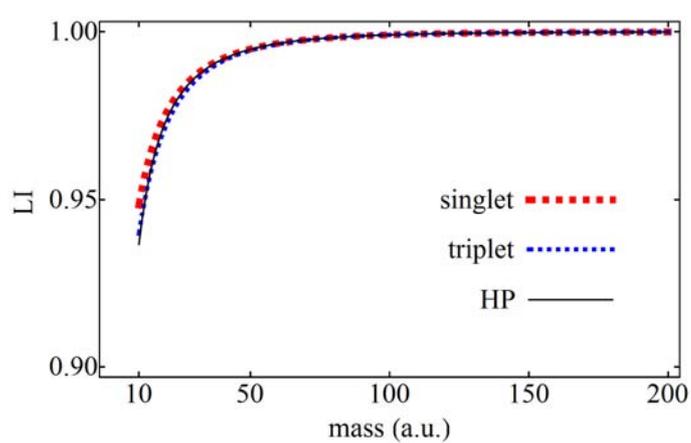

(d)

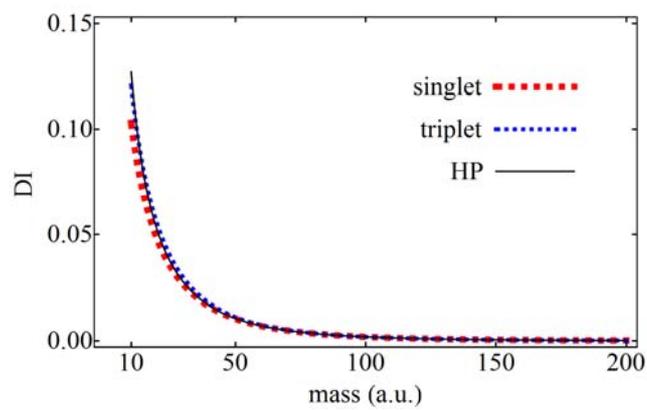

(e)

Figure-1

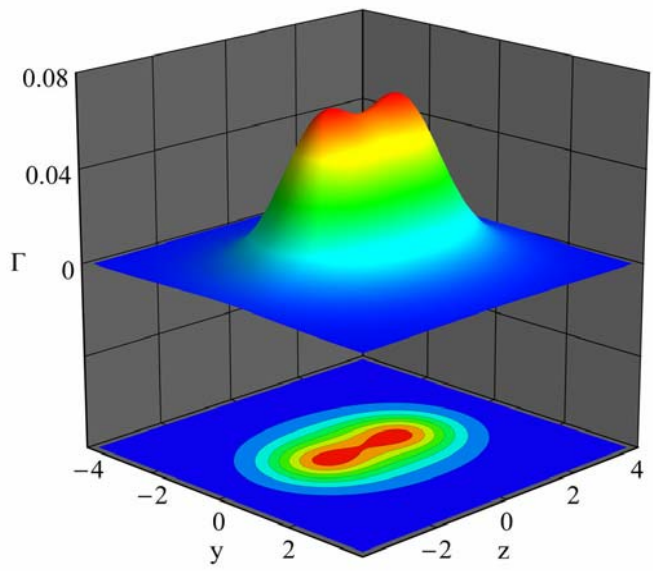
M=10m$_e$

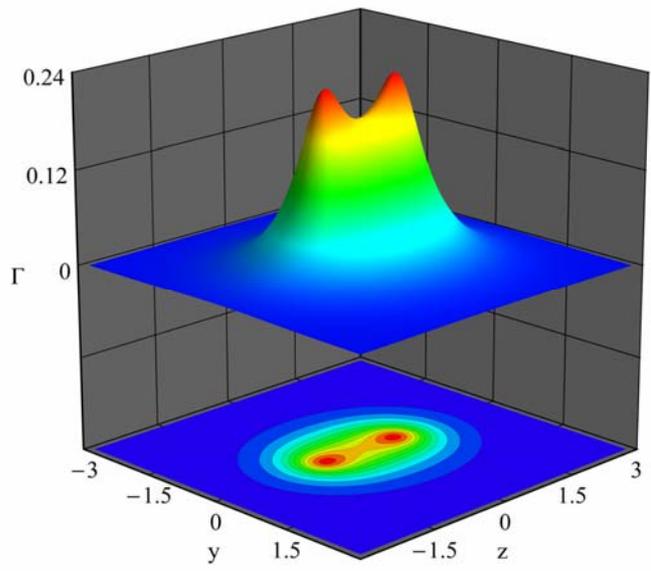
M=200m$_e$

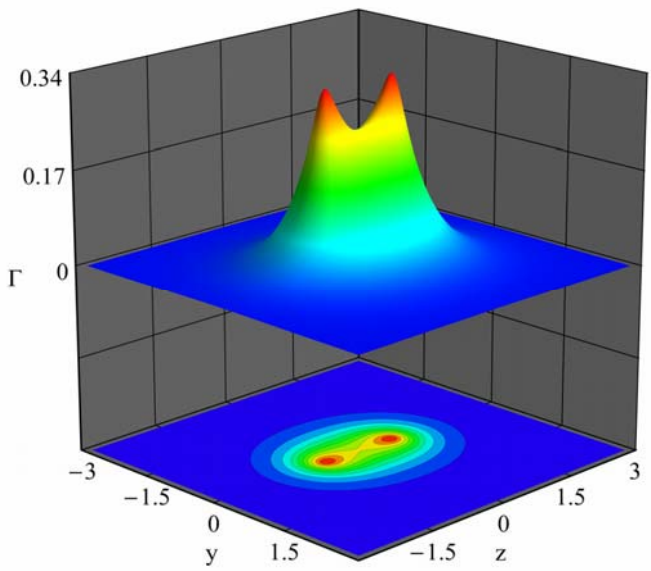
M=H

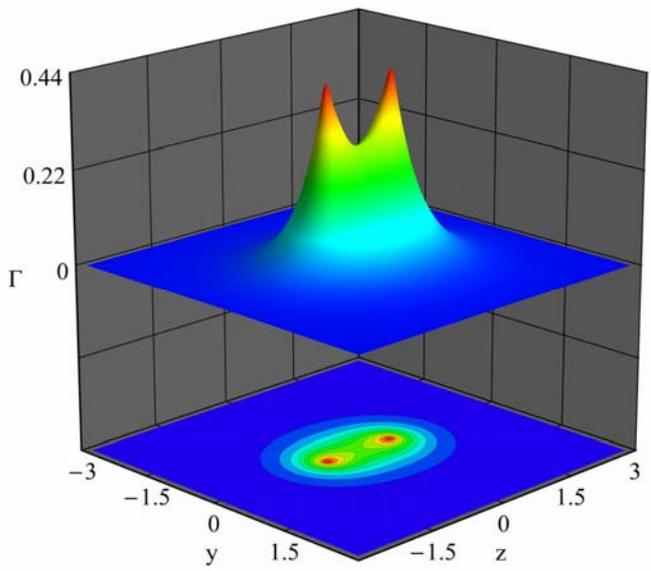
M=10$^{11}$m$_e$

Figure-2

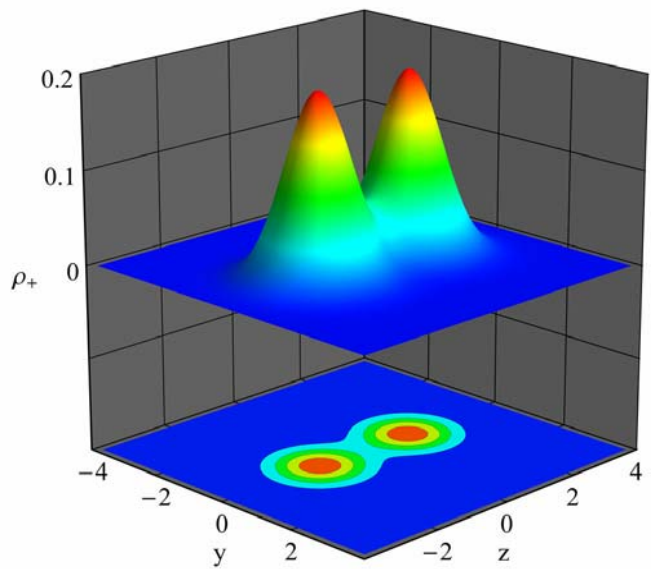

M=10m$_e$

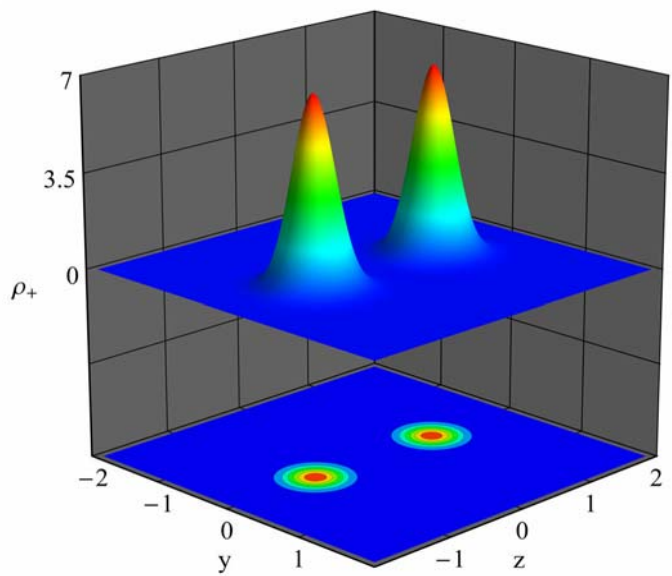

M=200m$_e$

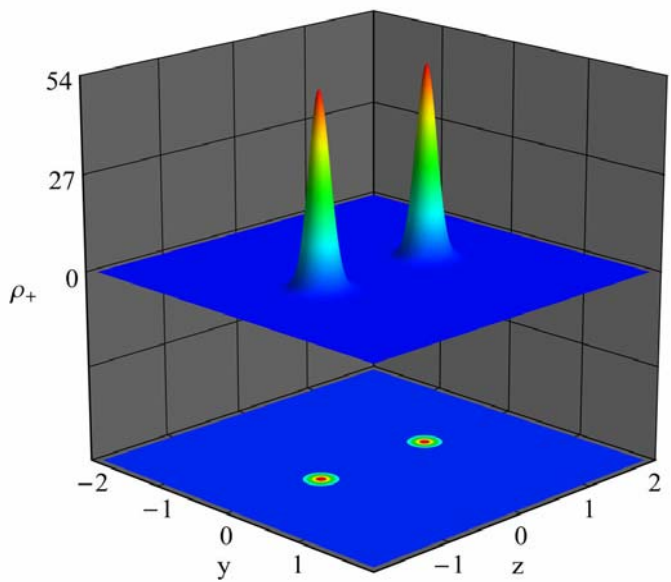

M=H

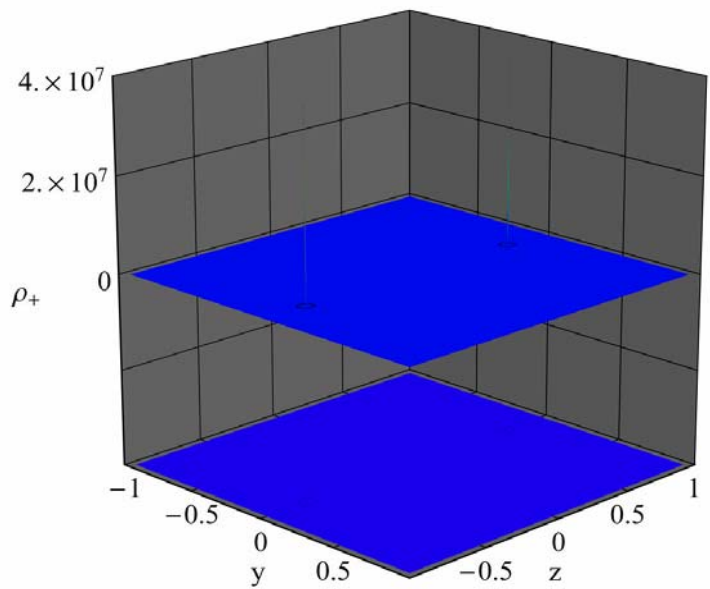

M=10$^{11}$m$_e$

Figure-3

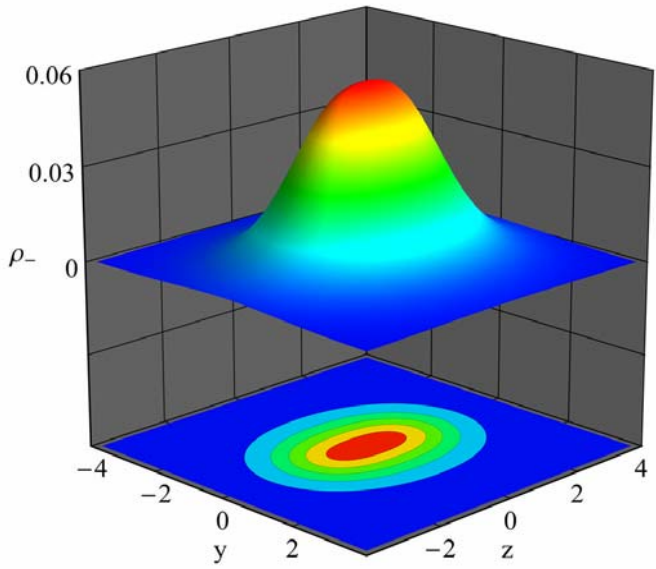

M=10m$_e$

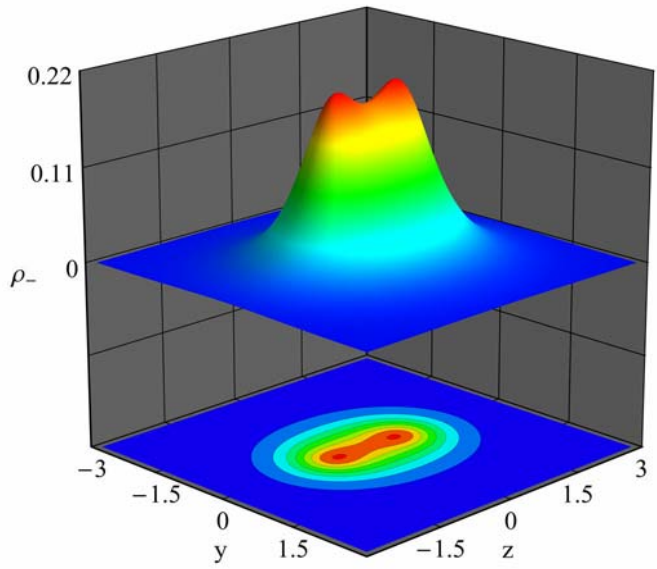

M=200m$_e$

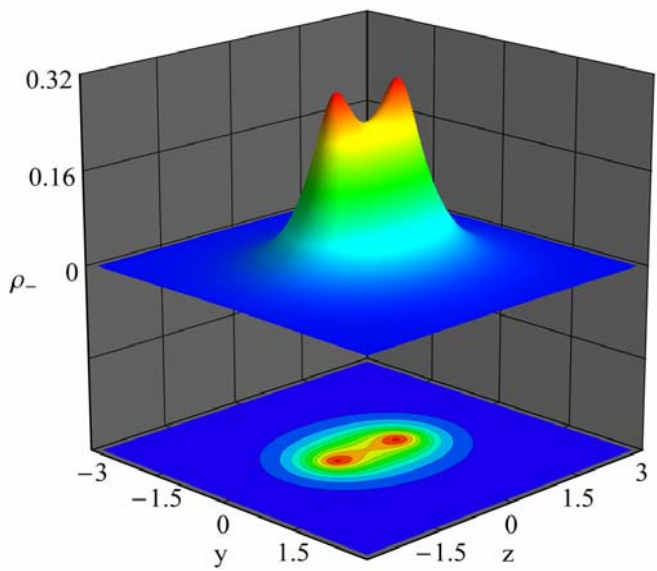

M=H

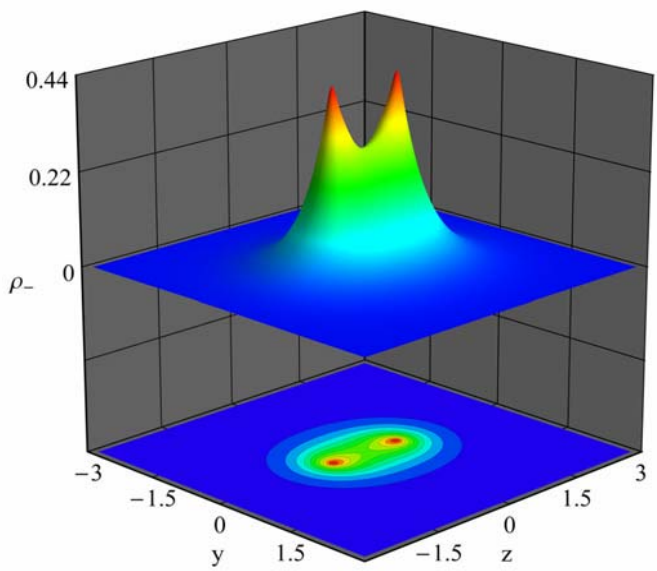

M=10$^{11}$m$_e$

Figure-4

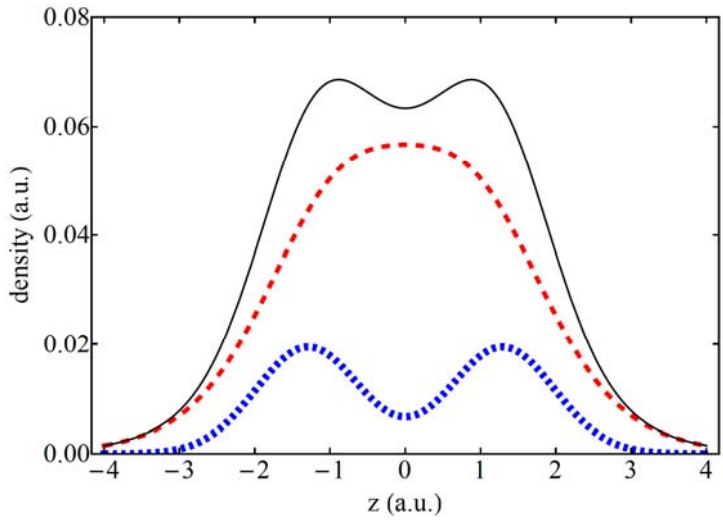

M=10m_e

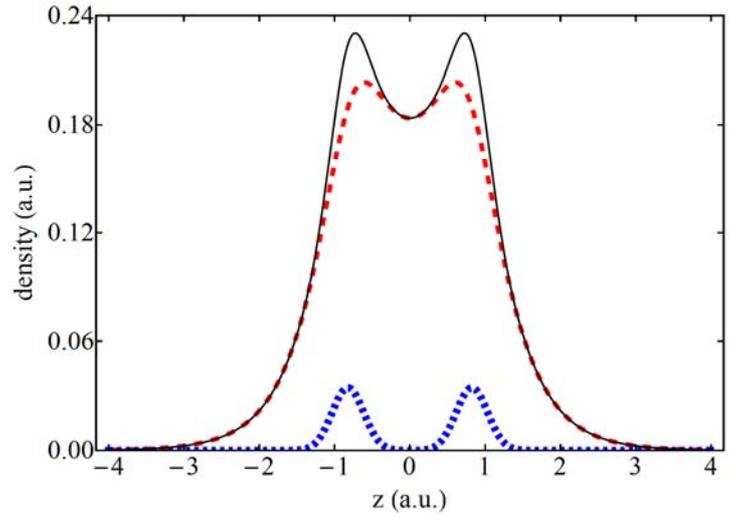

M=200m_e

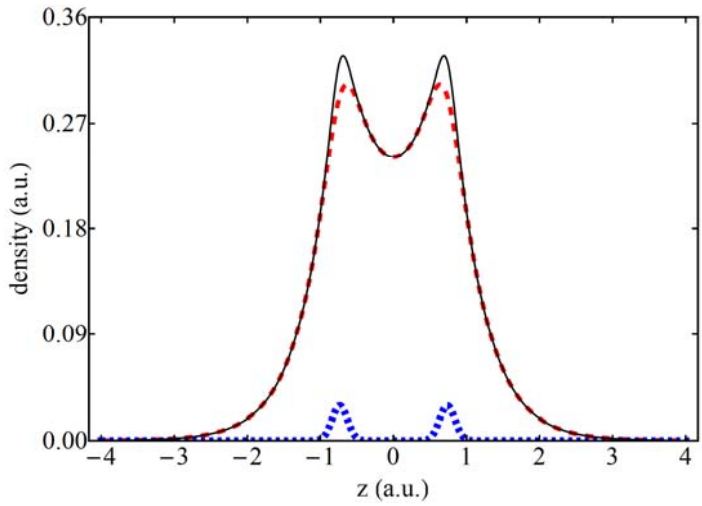

M=H

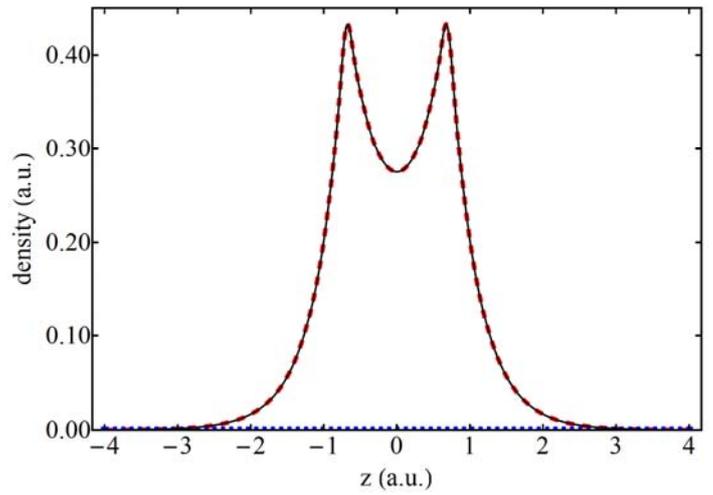

M=10$^{11}$m_e

Figure-5

Table 1- Some results of ab initio calculations at the FV-MC_MO/HP level employing [6s:1s] basis set.[*]

| Mass | energy | mean inter-particle distance[**] | exponent[***] | Energy components | | | | |
|---|---|---|---|---|---|---|---|---|
| | | | | kinetic-e | kinetic-PCP | e-e | PCP-PCP | e-PCP |
| *10* | -0.540577 | 2.5812 | 0.5274 | 0.382349 | 0.158230 | 0.399215 | 0.384316 | -1.864688 |
| *50* | -0.775669 | 1.9198 | 2.1165 | 0.648678 | 0.126989 | 0.514340 | 0.520867 | -2.586543 |
| *100* | -0.856426 | 1.7578 | 3.5298 | 0.750537 | 0.105893 | 0.550914 | 0.568903 | -2.832673 |
| *200* | -0.922596 | 1.6462 | 5.6865 | 0.837298 | 0.085297 | 0.579513 | 0.607455 | -3.032160 |
| *600* | -1.000365 | 1.5352 | 11.4915 | 0.942907 | 0.057458 | 0.611454 | 0.651343 | -3.263527 |
| *1000* | -1.026981 | 1.5016 | 15.6731 | 0.979962 | 0.047019 | 0.621949 | 0.665940 | -3.341851 |
| *1400* | -1.041776 | 1.4836 | 19.1477 | 1.000745 | 0.041031 | 0.627665 | 0.673992 | -3.385209 |
| *H* | -1.052294 | 1.4714 | 22.4286 | 1.015649 | 0.036645 | 0.631708 | 0.679591 | -3.415886 |
| *D* | -1.074311 | 1.4466 | 33.3807 | 1.047027 | 0.027283 | 0.639983 | 0.691311 | -3.479915 |
| *T* | -1.084341 | 1.4358 | 41.8364 | 1.061511 | 0.022833 | 0.643741 | 0.696498 | -3.508923 |
| $10^4$ | -1.096364 | 1.4230 | 58.1897 | 1.078908 | 0.017457 | 0.648140 | 0.702738 | -3.543606 |
| $10^5$ | -1.121095 | 1.3982 | 198.4822 | 1.115141 | 0.005954 | 0.657002 | 0.715167 | -3.614359 |
| $10^6$ | -1.129546 | 1.3900 | 654.3532 | 1.127580 | 0.001963 | 0.659862 | 0.719450 | -3.638401 |
| $10^7$ | -1.132252 | 1.3874 | 2103.369 | 1.131618 | 0.000631 | 0.660790 | 0.720736 | -3.646028 |
| $10^8$ | -1.133118 | 1.3866 | 6691.483 | 1.132914 | 0.000201 | 0.661086 | 0.721145 | -3.648464 |
| $10^9$ | -1.133392 | 1.3864 | 21196.5 | 1.133327 | 0.000064 | 0.661180 | 0.721275 | -3.649237 |
| $10^{10}$ | -1.133479 | 1.3864 | 67109.0 | 1.133459 | 0.000020 | 0.661210 | 0.721317 | -3.649486 |
| $10^{11}$ | -1.133507 | 1.3864 | 212098.8 | 1.133492 | 0.000006 | 0.661217 | 0.721330 | -3.649552 |
| $10^{12}$ | -1.133516 | 1.3864 | 647161.9 | 1.133512 | 0.000002 | 0.661222 | 0.721333 | -3.649585 |
| $10^{13}$ | -1.133518 | 1.3864 | 2114067.0 | 1.133516 | 0.000001 | 0.661223 | 0.721335 | -3.649593 |
| $\infty$ [****] | -1.133520 | 1.3864 | -- | 1.133509 | -- | 0.661221 | 0.721335 | -3.649584 |

[*] The symbol e in the energy component analysis is an abbreviation for electronic contribution.
[**] This is the distance between the maxima of the two Gaussian functions describing the PCPs.
[***] This is the exponent of the PCPs' Gaussian function, see equations (18) in the text.
[****] The calculation was done within clamped nucleus model thus only electrons were treated as quantum waves. In this case the basis set is composed of 10 s-type Gaussian functions describing just electrons. The inter-nuclear distance was determined by the orthodox geometry optimization.

Table 2- Some results of ab initio calculations at the FV-MC_MO/HP level employing [5s:1s1p1d] basis set.[*]

| mass | energy | Energy components | | | | | $\Delta E$[**] |
|---|---|---|---|---|---|---|---|
| | | kinetic-e | kinetic-PCP | e-e | PCP-PCP | e-PCP | |
| *10* | -0.546120 | 0.387969 | 0.158146 | 0.403801 | 0.389707 | -1.885743 | -0.005543 |
| *50* | -0.779073 | 0.653860 | 0.125216 | 0.517920 | 0.528330 | -2.604399 | -0.003404 |
| *100* | -0.858726 | 0.754213 | 0.104520 | 0.553413 | 0.574452 | -2.845325 | -0.002300 |
| *200* | -0.924098 | 0.839753 | 0.084356 | 0.581188 | 0.611309 | -3.040704 | -0.001503 |
| *600* | -1.001108 | 0.944139 | 0.056976 | 0.612322 | 0.653425 | -3.267970 | -0.000742 |
| *1000* | -1.027511 | 0.980846 | 0.046668 | 0.622589 | 0.667505 | -3.345119 | -0.000530 |
| *1400* | -1.042195 | 1.001471 | 0.040725 | 0.628204 | 0.675239 | -3.387834 | -0.000419 |
| *H* | -1.052644 | 1.016236 | 0.036405 | 0.632151 | 0.680685 | -3.418121 | -0.000351 |
| *D* | -1.074502 | 1.047388 | 0.027096 | 0.640295 | 0.691974 | -3.481256 | -0.000192 |
| *T* | -1.084541 | 1.061817 | 0.022722 | 0.643968 | 0.697047 | -3.510095 | -0.000200 |
| *$10^4$* | -1.096469 | 1.079083 | 0.017378 | 0.648298 | 0.703096 | -3.544323 | -0.000105 |
| *$10^5$* | -1.121114 | 1.115188 | 0.005925 | 0.657050 | 0.715360 | -3.614638 | -0.000019 |

[*] The symbol e in the energy component analysis is an abbreviation for electronic contribution.
[**] This is the difference of total energies in second column with those of Table 1: $\Delta E$ =E([5s:1s1p1d])-E([6s:1s]).

Table 3- Some results of ab initio calculations at the FV-MC_MO/HF (single/triplet) levels employing [6s:1s] basis set.*

*Singlet*

| Mass | energy | mean inter-particle distance** | exponent*** | coefficient*** | coefficient*** |
|---|---|---|---|---|---|
| 10 | -0.541120 | 2.5348 | 0.5198 | 1.004864 | -0.027821 |
| 50 | -0.775691 | 1.9186 | 2.1146 | 1.000059 | -0.003129 |
| 100 | -0.856427 | 1.7577 | 3.5295 | 1.000002 | -0.000584 |
| 200 | -0.922596 | 1.6462 | 5.6865 | 1.000000 | -0.000051 |
| 600 | -1.000365 | 1.5353 | 11.4917 | 1.000000 | 0.000000 |
| 1000 | -1.026981 | 1.5016 | 15.6734 | 1.000000 | 0.000000 |
| 1400 | -1.041773 | 1.4838 | 19.1388 | 1.000000 | 0.000000 |
| H | -1.052294 | 1.4715 | 22.4286 | 1.000000 | 0.000000 |

| | Energy components | | | | |
|---|---|---|---|---|---|
| | kinetic-e | kinetic- PCP | e-e | PCP - PCP | e- PCP |
| 10 | 0.383340 | 0.157780 | 0.400093 | 0.386193 | -1.868525 |
| 50 | 0.648780 | 0.126915 | 0.514405 | 0.521075 | -2.586866 |
| 100 | 0.750541 | 0.105887 | 0.550917 | 0.568915 | -2.832687 |
| 200 | 0.837298 | 0.085297 | 0.579513 | 0.607455 | -3.032160 |
| 600 | 0.942912 | 0.057458 | 0.611456 | 0.651346 | -3.263538 |
| 1000 | 0.979965 | 0.047020 | 0.621949 | 0.665942 | -3.341857 |
| 1400 | 1.000764 | 0.041012 | 0.627679 | 0.673947 | -3.385174 |
| H | 1.015649 | 0.036645 | 0.631708 | 0.679591 | -3.415886 |

*Triplet*

| Mass | Energy | mean inter-particle distance** | exponent*** | orbital-1 coefficient*** | orbital-1 coefficient*** | orbital-2 coefficient*** | orbital-2 coefficient*** |
|---|---|---|---|---|---|---|---|
| 10 | -0.543070 | 2.4220 | 0.5029 | 0.637892 | 0.637892 | 0.805187 | -0.805188 |
| 50 | -0.775896 | 1.9112 | 2.1027 | 0.699629 | 0.699629 | 0.714830 | -0.714830 |
| 100 | -0.856442 | 1.7571 | 3.5270 | 0.705591 | 0.705577 | 0.708632 | -0.708646 |
| 200 | -0.922596 | 1.6462 | 5.6863 | 0.706897 | 0.706999 | 0.707317 | -0.707215 |
| 600 | -1.000365 | 1.5353 | 11.4917 | 0.694325 | 0.719661 | 0.719662 | -0.694326 |
| 1000 | -1.026981 | 1.5016 | 15.6734 | 0.230897 | 0.972978 | 0.972978 | -0.230897 |
| 1400 | -1.041773 | 1.4838 | 19.1388 | 0.008870 | 0.999961 | 0.999961 | -0.008870 |
| H | -1.052294 | 1.4715 | 22.4286 | 0.000000 | 1.000000 | 1.000000 | 0.000000 |

| | Energy components | | | | |
|---|---|---|---|---|---|
| | kinetic-e | kinetic- PCP | e-e | PCP - PCP | e- PCP |
| 10 | 0.384018 | 0.159055 | 0.401034 | 0.384888 | -1.872065 |
| 50 | 0.649586 | 0.126310 | 0.514897 | 0.522679 | -2.589368 |
| 100 | 0.750627 | 0.105816 | 0.550967 | 0.569096 | -2.832948 |
| 200 | 0.837300 | 0.085295 | 0.579514 | 0.607461 | -3.032167 |
| 600 | 0.942912 | 0.057458 | 0.611456 | 0.651346 | -3.263538 |
| 1000 | 0.979965 | 0.047020 | 0.621949 | 0.665942 | -3.341857 |
| 1400 | 1.000764 | 0.041012 | 0.627679 | 0.673947 | -3.385174 |
| H | 1.015649 | 0.036645 | 0.631708 | 0.679591 | -3.415886 |

*The symbol e in the energy component analysis is an abbreviation for electronic contribution.
** This is the distance between the maxima of the two Gaussian functions describing the PCPs.
*** These are the exponent and coefficients of the Gaussian functions used to construct spatial orbitals of the PCPs. For a detailed discussion see the text above equations (22) and below equations (24).

Table 4- Some results of topological analysis of $\Gamma^{(2)}(\vec{q})$ employing the HP wavefunction with [6s:1s] basis set all offered in atomic units.*

| mass | CP type | am. of Gamma | Lap. Gamma | Hess.** | com.** | G | K | V | dis. from (3, -1) CP |
|---|---|---|---|---|---|---|---|---|---|
| *10* | **(3, -3)** | 0.0687 | -0.2249 | -0.0895 | -0.0458 | 0.0023 | 0.0585 | -0.0608 | 0.8824 |
| *50* | **(3, -3)** | 0.1524 | -1.4474 | -0.5386 | -0.3702 | 0.0089 | 0.3708 | -0.3797 | 0.7880 |
| *100* | **(3, -3)** | 0.1922 | -2.6012 | -0.9534 | -0.6944 | 0.0146 | 0.6649 | -0.6794 | 0.7506 |
| *200* | **(3, -3)** | 0.2303 | -4.2738 | -1.5465 | -1.1808 | 0.0220 | 1.0905 | -1.1125 | 0.7254 |
| *600* | **(3, -3)** | 0.2837 | -8.1879 | -2.9167 | -2.3546 | 0.0378 | 2.0848 | -2.1226 | 0.7022 |
| *1000* | **(3, -3)** | 0.3050 | -10.5930 | -3.7527 | -3.0876 | 0.0471 | 2.6954 | -2.7425 | 0.6958 |
| *1400* | **(3, -3)** | 0.3189 | -12.8621 | -4.5196 | -3.8229 | 0.0506 | 3.2662 | -3.3168 | 0.6943 |
| *H* | **(3, -3)** | 0.3272 | -13.9627 | -4.9188 | -4.1252 | 0.0602 | 3.5509 | -3.6111 | 0.6905 |
| *D* | **(3, -3)** | 0.3503 | -19.3622 | -6.7585 | -5.8452 | 0.0726 | 4.9131 | -4.9857 | 0.6882 |
| *T* | **(3, -3)** | 0.3591 | -21.5511 | -7.5281 | -6.4949 | 0.0875 | 5.4753 | -5.5628 | 0.6860 |
| *$10^4$* | **(3, -3)** | 0.3722 | -26.1147 | -9.1324 | -7.8498 | 0.1140 | 6.6427 | -6.7567 | 0.6837 |
| *$10^5$* | **(3, -3)** | 0.4044 | -49.4584 | -17.3079 | -14.8426 | 0.2530 | 12.6236 | -12.8825 | 0.6834 |
| *$10^6$* | **(3, -3)** | 0.4253 | -92.9190 | -32.4979 | -27.9232 | 0.5247 | 23.7545 | -24.2792 | 0.6861 |
| *$10^7$* | **(3, -3)** | 0.4308 | -143.0346 | -51.1004 | -40.8337 | 1.2396 | 36.9982 | -38.2378 | 0.6880 |
| *$10^8$* | **(3, -3)** | 0.4313 | -42.4065 | -16.8126 | -8.7812 | 0.8674 | 11.4690 | -11.4690 | 0.6748 |
| *$10^9$* | **(3, -3)** | 0.4324 | -48.3963 | -16.5012 | -15.3938 | 0.0005 | 12.0995 | -12.0995 | 0.6752 |
| *$10^{10}$* | **(3, -3)** | 0.4327 | -48.8480 | -16.6519 | -15.5443 | 0.0000 | 12.2120 | -12.2120 | 0.6753 |
| *$10^{11}$* | **(3, -3)** | 0.4327 | -48.6262 | -16.5762 | -15.4737 | 0.0000 | 12.1565 | -12.1565 | 0.6752 |
| *$10^{12}$* | **(3, -3)** | 0.4329 | -49.0358 | -16.7150 | -15.6059 | 0.0000 | 12.2589 | -12.2589 | 0.6753 |
| *$10^{13}$* | **(3, -3)** | 0.4329 | -49.0369 | -16.7152 | -15.6064 | 0.0000 | 12.2592 | -12.2592 | 0.6753 |
| ∞*** | **H** | 0.4327 | -48.6617 | -16.5880 | -15.4856 | 0.0000 | 12.1654 | -12.1654 | 0.6752 |
| *10* | **(3, -1)** | 0.0634 | -0.1114 | -0.0701 | 0.0289 | 0.0062 | 0.0341 | -0.0403 | |
| *50* | **(3, -1)** | 0.1271 | -0.3745 | -0.2499 | 0.1253 | 0.0105 | 0.1041 | -0.1145 | |
| *100* | **(3, -1)** | 0.1568 | -0.5651 | -0.3602 | 0.1553 | 0.0056 | 0.1468 | -0.1524 | |
| *200* | **(3, -1)** | 0.1838 | -0.7591 | -0.4783 | 0.1975 | 0.0014 | 0.1911 | -0.1925 | |
| *600* | **(3, -1)** | 0.2180 | -1.0265 | -0.6554 | 0.2842 | 0.0000 | 0.2567 | -0.2567 | |
| *1000* | **(3, -1)** | 0.2301 | -1.1288 | -0.7264 | 0.3240 | 0.0000 | 0.2822 | -0.2822 | |
| *1400* | **(3, -1)** | 0.2346 | -1.0737 | -0.7423 | 0.4110 | 0.0000 | 0.2684 | -0.2684 | |
| *H* | **(3, -1)** | 0.2416 | -1.2165 | -0.7951 | 0.3736 | 0.0000 | 0.3041 | -0.3041 | |
| *D* | **(3, -1)** | 0.2494 | -1.1688 | -0.8300 | 0.4912 | 0.0000 | 0.2922 | -0.2922 | |
| *T* | **(3, -1)** | 0.2531 | -1.1830 | -0.8596 | 0.5363 | 0.0000 | 0.2957 | -0.2957 | |
| *$10^4$* | **(3, -1)** | 0.2615 | -1.3269 | -0.9288 | 0.5306 | 0.0000 | 0.3317 | -0.3317 | |
| *$10^5$* | **(3, -1)** | 0.2698 | -1.2405 | -0.9562 | 0.6719 | 0.0000 | 0.3101 | -0.3101 | |
| *$10^6$* | **(3, -1)** | 0.2743 | -1.4166 | -0.9958 | 0.5750 | 0.0000 | 0.3541 | -0.3541 | |
| *$10^7$* | **(3, -1)** | 0.2751 | -1.3896 | -1.0025 | 0.6154 | 0.0000 | 0.3474 | -0.3474 | |
| *$10^8$* | **(3, -1)** | 0.2753 | -1.3768 | -1.0041 | 0.6313 | 0.0000 | 0.3442 | -0.3442 | |
| *$10^9$* | **(3, -1)** | 0.2754 | -1.3725 | -1.0045 | 0.6365 | 0.0000 | 0.3431 | -0.3431 | |
| *$10^{10}$* | **(3, -1)** | 0.2754 | -1.3711 | -1.0047 | 0.6383 | 0.0000 | 0.3428 | -0.3428 | |
| *$10^{11}$* | **(3, -1)** | 0.2755 | -1.3746 | -1.0050 | 0.6355 | 0.0000 | 0.3436 | -0.3436 | |
| *$10^{12}$* | **(3, -1)** | 0.2754 | -1.3705 | -1.0048 | 0.6390 | 0.0000 | 0.3426 | -0.3426 | |
| *$10^{13}$* | **(3, -1)** | 0.2754 | -1.3707 | -1.0048 | 0.6389 | 0.0000 | 0.3427 | -0.3427 | |
| ∞*** | **BCP** | 0.2755 | -1.3747 | -1.0051 | 0.6355 | 0.0000 | 0.3437 | -0.3437 | |

* The abbreviations am., Lap., Hess., com. and dis. are used for amount, Laplacian, Hessian, components and distance respectively. The symbol ∞ is used for entry of the orthodox QTAIM analysis based on the clamped nucleus model.
** The first column contains the degenerate Hessian component.
*** The calculation was done within clamped nucleus model thus only electrons were treated as quantum waves. In this case the basis set is composed of 10 s-type Gaussian functions describing just electrons.

Table 5- Some results of topological analysis of $\rho_+(\vec{q})$ employing the HP wavefunction with [6s:1s] basis set all offered in atomic units.*

| Mass | CP type | am. of $\rho_+$ | Lap. $\rho_+$ | Hess.** | com.** | dis. from (3, -1) CP | K | V |
|---|---|---|---|---|---|---|---|---|
| 10 | (3, -3) | 0.1946 | -1.2274 | -0.4109 | -0.4057 | 1.2883 | 0.0307 | -0.0308 |
| 50 | (3, -3) | 1.5640 | -39.7227 | -13.2409 | -13.2408 | 0.9599 | 0.1986 | -0.1986 |
| 100 | (3, -3) | 3.3685 | -142.6821 | -47.5607 | -47.5607 | 0.8789 | 0.3567 | -0.3567 |
| 200 | (3, -3) | 6.8879 | -470.0169 | -156.6723 | -156.6723 | 0.8231 | 0.5875 | -0.5875 |
| 600 | (3, -3) | 19.7873 | -2728.6336 | -909.5445 | -909.5445 | 0.7676 | 1.1369 | -1.1369 |
| 1000 | (3, -3) | 31.5176 | -5927.7443 | -1975.9148 | -1975.9148 | 0.7508 | 1.4819 | -1.4819 |
| 1400 | (3, -3) | 42.5595 | -9779.0040 | -3259.6680 | -3259.6680 | 0.7418 | 1.7463 | -1.7463 |
| H | (3, -3) | 53.9541 | -14521.4139 | -4840.4713 | -4840.4713 | 0.7357 | 1.9772 | -1.9772 |
| D | (3, -3) | 97.9633 | -39240.971 | -13080.324 | -13080.324 | 0.7233 | 2.6727 | -2.6727 |
| T | (3, -3) | 137.4522 | -69006.051 | -23002.017 | -23002.017 | 0.7179 | 3.1384 | -3.1384 |
| $10^4$ | (3, -3) | 225.4704 | -157440.690 | -52480.230 | -52480.230 | 0.7115 | 3.9360 | -3.9360 |
| $10^5$ | (3, -3) | 1420.3731 | -3383025.23 | -1127675.08 | -1127675.08 | 0.6991 | 8.4576 | -8.4576 |
| $10^6$ | (3, -3) | 8502.3407 | -66762406.2 | -22254135.4 | -22254135.4 | 0.6950 | 16.69060 | -16.69060 |
| $10^7$ | (3, -3) | 48999.6785 | -1236772857.0 | -412257619.0 | -412257619.0 | 0.6937 | 30.919321 | -30.919321 |
| | | | | | | G | | |
| 10 | (3, -1) | 0.0672 | 0.0728 | -0.1417 | 0.3562 | 0.0062 | 0.0044 | -0.0106 |
| 50 | (3, -1) | 0.0633 | 2.5730 | -0.5360 | 3.6451 | 0.0105 | -0.0024 | -0.0080 |
| 100 | (3, -1) | 0.0289 | 3.2212 | -0.4074 | 4.0360 | 0.0056 | -0.0025 | -0.0031 |
| 200 | (3, -1) | 0.0062 | 1.7517 | -0.1411 | 2.0340 | 0.0014 | -0.0008 | -0.0005 |
| 600 | (3, -1) | 0.0001 | 0.0575 | -0.0024 | 0.0623 | 0.0000 | 0.0000 | 0.0000 |
| 1000 | (3, -1) | 0.0000 | 0.0027 | -0.0001 | 0.0029 | 0.0000 | 0.0000 | 0.0000 |
| 1400 | (3, -1) | 0.0000 | 0.0002 | 0.0000 | 0.0002 | 0.0000 | 0.0000 | 0.0000 |
| H | (3, -1) | 0.0000 | 0.0000 | 0.0000 | 0.0000 | 0.0000 | 0.0000 | 0.0000 |
| D | (3, -1) | 0.0000 | 0.0000 | 0.0000 | 0.0000 | 0.0000 | 0.0000 | 0.0000 |
| T | (3, -1) | 0.0000 | 0.0000 | 0.0000 | 0.0000 | 0.0000 | 0.0000 | 0.0000 |
| $10^4$ | (3, -1) | 0.0000 | 0.0000 | 0.0000 | 0.0000 | 0.0000 | 0.0000 | 0.0000 |
| $10^5$ | (3, -1) | 0.0000 | 0.0000 | 0.0000 | 0.0000 | 0.0000 | 0.0000 | 0.0000 |
| $10^6$ | (3, -1) | 0.0000 | 0.0000 | 0.0000 | 0.0000 | 0.0000 | 0.0000 | 0.0000 |
| $10^7$ | (3, -1) | 0.0000 | 0.0000 | 0.0000 | 0.0000 | 0.0000 | 0.0000 | 0.0000 |

* The abbreviations am., Lap., Hess., com. and dis. are used for amount, Laplacian, Hessian, components and distance, respectively.
** The first column contains the degenerate Hessian component.

Table 6- Some results of topological analysis of $\rho_-(\vec{q})$ employing the HP wavefunction with [6s:1s] basis set all offered in atomic units.*

| mass | CP type | am. of $\rho_-$ | Lap. $\rho_-$ | Hess.** | com.** | dis. from (3, -1) CP | K | V |
|---|---|---|---|---|---|---|---|---|
| 10 | (3, -3) | 0.0567 | -0.1187 | -0.0560 | -0.0067 | -- | 0.0297 | -0.0297 |
| 50 | (3, -3) | 0.1294 | -0.6793 | -0.2897 | -0.0999 | 0.5293 | 0.1698 | -0.1698 |
| 100 | (3, -3) | 0.1665 | -1.2493 | -0.5044 | -0.2404 | 0.5787 | 0.3123 | -0.3123 |
| 200 | (3, -3) | 0.2034 | -2.0784 | -0.8064 | -0.4656 | 0.6035 | 0.5196 | -0.5196 |
| 600 | (3, -3) | 0.2572 | -3.9942 | -1.4836 | -1.0270 | 0.6246 | 0.9986 | -0.9986 |
| 1000 | (3, -3) | 0.2794 | -5.1497 | -1.8855 | -1.3786 | 0.6311 | 1.2874 | -1.2874 |
| 1400 | (3, -3) | 0.2937 | -6.3309 | -2.2915 | -1.7478 | 0.6416 | 1.5827 | -1.5827 |
| H | (3, -3) | 0.3032 | -6.7392 | -2.4316 | -1.8759 | 0.6372 | 1.6848 | -1.6848 |
| D | (3, -3) | 0.3279 | -9.4448 | -3.3595 | -2.7258 | 0.6490 | 2.3612 | -2.3612 |
| T | (3, -3) | 0.3385 | -10.0919 | -3.5733 | -2.9453 | 0.6471 | 2.5230 | -2.5230 |
| $10^4$ | (3, -3) | 0.3539 | -11.9706 | -4.2185 | -3.5335 | 0.6483 | 2.9926 | -2.9926 |
| $10^5$ | (3, -3) | 0.3936 | -19.9556 | -6.9210 | -6.1137 | 0.6584 | 4.9889 | -4.9889 |
| $10^6$ | (3, -3) | 0.4189 | -35.4786 | -12.1532 | -11.1722 | 0.6709 | 8.8697 | -8.8697 |
| $10^7$ | (3, -3) | 0.4279 | -43.4352 | -14.8317 | -13.7719 | 0.6738 | 10.8588 | -10.8588 |
| $10^{13}$ | (3, -3) | 0.4329 | -49.0369 | -16.7152 | -15.6064 | 0.6753 | 12.2592 | -12.2592 |
| ∞ | H | 0.4327 | -48.6617 | -16.5880 | -15.4856 | 0.6752 | 12.1654 | -12.1654 |
| | | | | | | G | | |
| 10 | (3, -1) | -- | -- | -- | -- | -- | -- | -- |
| 50 | (3, -1) | 0.1258 | -0.4260 | -0.2392 | 0.0524 | 0.0000 | 0.1065 | -0.1065 |
| 100 | (3, -1) | 0.1565 | -0.5973 | -0.3561 | 0.1149 | 0.0000 | 0.1493 | -0.1493 |
| 200 | (3, -1) | 0.1837 | -0.7678 | -0.4776 | 0.1873 | 0.0000 | 0.1920 | -0.1920 |
| 600 | (3, -1) | 0.2180 | -1.0266 | -0.6554 | 0.2841 | 0.0000 | 0.2567 | -0.2567 |
| 1000 | (3, -1) | 0.2301 | -1.1288 | -0.7264 | 0.3240 | 0.0000 | 0.2822 | -0.2822 |
| 1400 | (3, -1) | 0.2346 | -1.0737 | -0.7423 | 0.4110 | 0.0000 | 0.2684 | -0.2684 |
| H | (3, -1) | 0.2416 | -1.2165 | -0.7951 | 0.3736 | 0.0000 | 0.3041 | -0.3041 |
| D | (3, -1) | 0.2494 | -1.1688 | -0.8300 | 0.4912 | 0.0000 | 0.2922 | -0.2922 |
| T | (3, -1) | 0.2531 | -1.1830 | -0.8596 | 0.5363 | 0.0000 | 0.2957 | -0.2957 |
| $10^4$ | (3, -1) | 0.2615 | -1.3269 | -0.9288 | 0.5306 | 0.0000 | 0.3317 | -0.3317 |
| $10^5$ | (3, -1) | 0.2698 | -1.2405 | -0.9562 | 0.6719 | 0.0000 | 0.3101 | -0.3101 |
| $10^6$ | (3, -1) | 0.2743 | -1.4166 | -0.9958 | 0.5750 | 0.0000 | 0.3541 | -0.3541 |
| $10^7$ | (3, -1) | 0.2751 | -1.3896 | -1.0025 | 0.6154 | 0.0000 | 0.3474 | -0.3474 |
| $10^{13}$ | (3, -1) | 0.2754 | -1.3707 | -1.0048 | 0.6389 | 0.0000 | 0.3427 | -0.3427 |
| ∞ | BCP | 0.2755 | -1.3747 | -1.0051 | 0.6355 | 0.0000 | 0.3437 | -0.3437 |

* The abbreviations am., Lap., Hess., com. and dis. are used for amount, Laplacian, Hessian, components and distance, respectively.
** The first column contains the degenerate Hessian component.

Table 7- Some results of topological analysis of $\Gamma^{(2)}(\vec{q})$ employing the HP wavefunction with [5s:1s1p1d] basis set as well as singlet and triple states with [6s:1s] basis set all offered in atomic units.[*]

**HP-[5s:1s1p1d]**

| mass | CP type | am. of Gamma | Lap. Gamma | Hess.[**] | com.[**] | dis. from (3, -1) CP | K | V |
|---|---|---|---|---|---|---|---|---|
| 10  | (3, -3) | 0.0722 | -0.2645 | -0.0962 | -0.0721 | 0.8795 | 0.0686 | -0.0711 |
| 50  | (3, -3) | 0.1556 | -1.5378 | -0.5416 | -0.4545 | 0.7645 | 0.3949 | -0.4053 |
| 100 | (3, -3) | 0.1946 | -2.7188 | -0.9475 | -0.8238 | 0.7370 | 0.6962 | -0.7126 |
| 200 | (3, -3) | 0.2319 | -4.4229 | -1.5272 | -1.3685 | 0.7183 | 1.1299 | -1.1541 |
| 600 | (3, -3) | 0.2843 | -8.3766 | -2.8606 | -2.6554 | 0.7002 | 2.1346 | -2.1751 |
|     |         |        |         |         |         | G      |        |         |
| 10  | (3, -1) | 0.0637 | -0.0870 | -0.0669 | 0.0468  | 0.0092 | 0.0310 | -0.0402 |
| 50  | (3, -1) | 0.1294 | -0.3859 | -0.2487 | 0.1114  | 0.0081 | 0.1046 | -0.1126 |
| 100 | (3, -1) | 0.1589 | -0.5842 | -0.3619 | 0.1396  | 0.0023 | 0.1484 | -0.1507 |
| 200 | (3, -1) | 0.1851 | -0.7589 | -0.4795 | 0.2001  | 0.0002 | 0.1899 | -0.1902 |
| 600 | (3, -1) | 0.2179 | -0.9948 | -0.6484 | 0.3020  | 0.0000 | 0.2487 | -0.2487 |

**singlet-[6s:1s]**

| mass | CP type | am. of Gamma | Lap. Gamma | Hess.[**] | com.[**] | dis. from (3, -1) CP | K | V |
|---|---|---|---|---|---|---|---|---|
| 10  | (3, -3) | 0.0689 | -0.2219 | -0.0887 | -0.0444 | 0.8518 | 0.0579 | -0.0604 |
| 50  | (3, -3) | 0.1524 | -1.4454 | -0.5380 | -0.3694 | 0.7871 | 0.3703 | -0.3792 |
| 100 | (3, -3) | 0.1922 | -2.6008 | -0.9532 | -0.6943 | 0.7506 | 0.6648 | -0.6793 |
| 200 | (3, -3) | 0.2303 | -4.2737 | -1.5465 | -1.1808 | 0.7254 | 1.0905 | -1.1125 |
| 600 | (3, -3) | 0.2837 | -8.1900 | -2.9173 | -2.3553 | 0.7021 | 2.0853 | -2.1232 |
|     |         |        |         |         |         | G      |        |         |
| 10  | (3, -1) | 0.0641 | -0.1137 | -0.0710 | 0.0282  | 0.0066 | 0.0351 | -0.0417 |
| 50  | (3, -1) | 0.1271 | -0.3748 | -0.2501 | 0.1254  | 0.0106 | 0.1043 | -0.1148 |
| 100 | (3, -1) | 0.1568 | -0.5651 | -0.3602 | 0.1553  | 0.0056 | 0.1468 | -0.1524 |
| 200 | (3, -1) | 0.1838 | -0.7591 | -0.4783 | 0.1975  | 0.0014 | 0.1911 | -0.1925 |
| 600 | (3, -1) | 0.2180 | -1.0267 | -0.6553 | 0.2838  | 0.0000 | 0.2567 | -0.2567 |

**triplet-[6s:1s]**

| mass | CP type | am. of Gamma | Lap. Gamma | Hess.[**] | com.[**] | dis. from (3, -1) CP | K | V |
|---|---|---|---|---|---|---|---|---|
| 10  | (3, -3) | 0.0691 | -0.2181 | -0.0867 | -0.0446 | 0.8135 | 0.0573 | -0.0601 |
| 50  | (3, -3) | 0.1525 | -1.4325 | -0.5342 | -0.3640 | 0.7807 | 0.3671 | -0.3761 |
| 100 | (3, -3) | 0.1922 | -2.5976 | -0.9523 | -0.6931 | 0.7500 | 0.6640 | -0.6786 |
| 200 | (3, -3) | 0.2303 | -4.2734 | -1.5464 | -1.1807 | 0.7254 | 1.0904 | -1.1124 |
| 600 | (3, -3) | 0.2837 | -8.1900 | -2.9173 | -2.3553 | 0.7021 | 2.0853 | -2.1232 |
|     |         |        |         |         |         | G      |        |         |
| 10  | (3, -1) | 0.0646 | -0.1124 | -0.0707 | 0.0291  | 0.0080 | 0.0361 | -0.0440 |
| 50  | (3, -1) | 0.1278 | -0.3783 | -0.2521 | 0.1260  | 0.0110 | 0.1056 | -0.1165 |
| 100 | (3, -1) | 0.1568 | -0.5656 | -0.3605 | 0.1554  | 0.0056 | 0.1470 | -0.1526 |
| 200 | (3, -1) | 0.1838 | -0.7591 | -0.4783 | 0.1975  | 0.0014 | 0.1911 | -0.1925 |
| 600 | (3, -1) | 0.2180 | -1.0267 | -0.6553 | 0.2838  | 0.0000 | 0.2567 | -0.2567 |

[*] The abbreviations am., Lap., Hess., com. and dis. are used for amount, Laplacian, Hessian, components and distance, respectively.
[**] The first column contains the degenerate Hessian component.

Table 8- Some results of basin integrations employing the HP wavefunction with [6s:1s] basis set all offered in atomic units.

| mass | Energies | | | | Electronic polarization dipole | Atomic volume | ESD |
|---|---|---|---|---|---|---|---|
| | Basin | electronic | PCP | $\Delta E^*$ | | | |
| *10* | -0.27029 | -0.19118 | -0.07912 | 0.00001 | -- | 126.27 | 3.6 |
| *50* | -0.38784 | -0.32432 | -0.06350 | 0.00001 | 0.058 | 86.67 | 1.8 |
| *100* | -0.42822 | -0.37526 | -0.05295 | 0.00001 | 0.068 | 78.28 | 1.4 |
| *200* | -0.46130 | -0.41865 | -0.04265 | 0.00001 | 0.078 | 72.66 | 1.1 |
| *600* | -0.50019 | -0.47145 | -0.02873 | 0.00001 | 0.086 | 67.15 | 0.8 |
| *1000* | -0.51350 | -0.48998 | -0.02352 | 0.00001 | 0.089 | 65.47 | 0.7 |
| *1400* | -0.52089 | -0.50038 | -0.02052 | 0.00001 | 0.090 | 64.53 | 0.6 |
| *H* | -0.52615 | -0.50783 | -0.01832 | 0.00000 | 0.091 | 63.94 | 0.6 |
| *D* | -0.53716 | -0.52352 | -0.01364 | 0.00000 | 0.093 | 62.73 | 0.5 |
| *T* | -0.54217 | -0.53076 | -0.01142 | 0.00000 | 0.094 | 62.04 | 0.5 |
| $10^4$ | -0.54818 | -0.53946 | -0.00873 | 0.00001 | 0.095 | 61.44 | 0.4 |
| $10^5$ | -0.56055 | -0.55757 | -0.00298 | 0.00001 | 0.096 | 60.08 | 0.2 |
| $10^6$ | -0.56478 | -0.56379 | -0.00098 | 0.00001 | 0.097 | 59.92 | 0.2 |
| $10^7$ | -0.56613 | -0.56581 | -0.00032 | 0.00001 | 0.097 | 59.74 | 0.1 |
| $10^8$ | -0.56656 | -0.56646 | -0.00010 | 0.00000 | 0.097 | 59.68 | 0.0 |
| $10^9$ | -0.56670 | -0.56667 | -0.00003 | 0.00000 | 0.097 | 59.66 | 0.0 |
| $10^{10}$ | -0.56674 | -0.56673 | -0.00001 | 0.00001 | 0.097 | 59.66 | 0.0 |
| $10^{11}$ | -0.56676 | -0.56675 | 0.00000 | 0.00001 | 0.097 | 59.66 | 0.0 |
| $10^{12}$ | -0.56676 | -0.56676 | 0.00000 | 0.00000 | 0.097 | 59.65 | 0.0 |
| $10^{13}$ | -0.56676 | -0.56676 | 0.00000 | 0.00000 | 0.097 | 59.65 | 0.0 |
| $\infty^{**}$ | -0.56676 | -- | -- | -- | 0.097 | 59.66 | -- |

* $\Delta E = E(\text{ab initio}) - 2E(\text{Basin})$
** The calculation was done within the clamped nucleus model thus only electrons were treated as quantum waves. In this case the basis set is composed of 10 s-type Gaussian functions describing just electrons. The computed wavefunction was then used for the orthodox QTAIM analysis.

Table 9- Some results of basin integrations employing the HP wavefunction with [5s:1s1p1d] basis set as well as singlet and triple states with [6s:1s] basis set all offered in atomic units.

| HP-[5s:1s1p1d] mass | Basin energy | ΔE* | Electronic polarization dipole | Atomic volume | PCP polarization dipole |
|---|---|---|---|---|---|
| *10* | -0.27306 | 0.00001 | -- | 124.55 | 0.059 |
| *50* | -0.38954 | 0.00001 | 0.091 | 85.91 | 0.022 |
| *100* | -0.42936 | 0.00000 | 0.092 | 77.83 | 0.015 |
| *200* | -0.46205 | 0.00001 | 0.093 | 72.38 | 0.010 |
| *600* | -0.50056 | 0.00001 | 0.094 | 67.01 | 0.005 |
| **singlet** | | | | | |
| *10* | -0.27057 | 0.00001 | -- | 125.94 | |
| *50* | -0.38785 | 0.00001 | 0.058 | 86.66 | |
| *100* | -0.42822 | 0.00001 | 0.068 | 78.28 | |
| *200* | -0.46130 | 0.00001 | 0.078 | 72.66 | |
| *600* | -0.50018 | 0.00000 | 0.086 | 67.14 | |
| **triplet** | | | | | |
| *10* | -0.27154 | 0.00001 | -- | 125.63 | |
| *50* | -0.38795 | 0.00001 | 0.057 | 86.55 | |
| *100* | -0.42823 | 0.00001 | 0.068 | 78.27 | |
| *200* | -0.46130 | 0.00001 | 0.078 | 72.66 | |
| *600* | -0.50019 | 0.00001 | 0.086 | 67.14 | |

* ΔE=E(ab initio)-2E(Basin)

Table 10- The computed localization and delocalization indexes for the PCPs employing all considered wavefunctions.

| Mass | HP-[6s:1s] LI | HP-[6s:1s] DI | HP-[5s:1s1p1d] LI | HP-[5s:1s1p1d] DI | singlet LI | singlet DI | triplet LI | triplet DI |
|---|---|---|---|---|---|---|---|---|
| *10* | 0.9410 | 0.1180 | 0.9557 | 0.0887 | 0.9425 | 0.1150 | 0.9409 | 0.1182 |
| *50* | 0.9948 | 0.0104 | 0.9979 | 0.0043 | 0.9948 | 0.0103 | 0.9947 | 0.0107 |
| *100* | 0.9990 | 0.0019 | 0.9997 | 0.0005 | 0.9990 | 0.0019 | 0.9990 | 0.0019 |
| *200* | 0.9999 | 0.0002 | 1.0000 | 0.0000 | 0.9999 | 0.0002 | 0.9999 | 0.0002 |
| *600* | 1.0000 | 0.0000 | 1.0000 | 0.0000 | 1.0000 | 0.0000 | 1.0000 | 0.0000 |

# Supporting Information

The Two-Component Quantum Theory of Atoms in Molecules (TC-QTAIM): The unified theory of localization/delocalization of electrons, nuclei and exotic elementary particles

Mohammad Goli and Shant Shahbazian[*]

Department of Chemistry, Faculty of Sciences, Shahid Beheshti University, G. C., Evin, Tehran, Iran, 19839, P.O. Box 19395-4716.

Tel/Fax: 98-21-22431661

E-mails:

(Shant Shahbazian) chemist_shant@yahoo.com

(Mohammad Goli) muhammadgoli@yahoo.com

# Table of contents

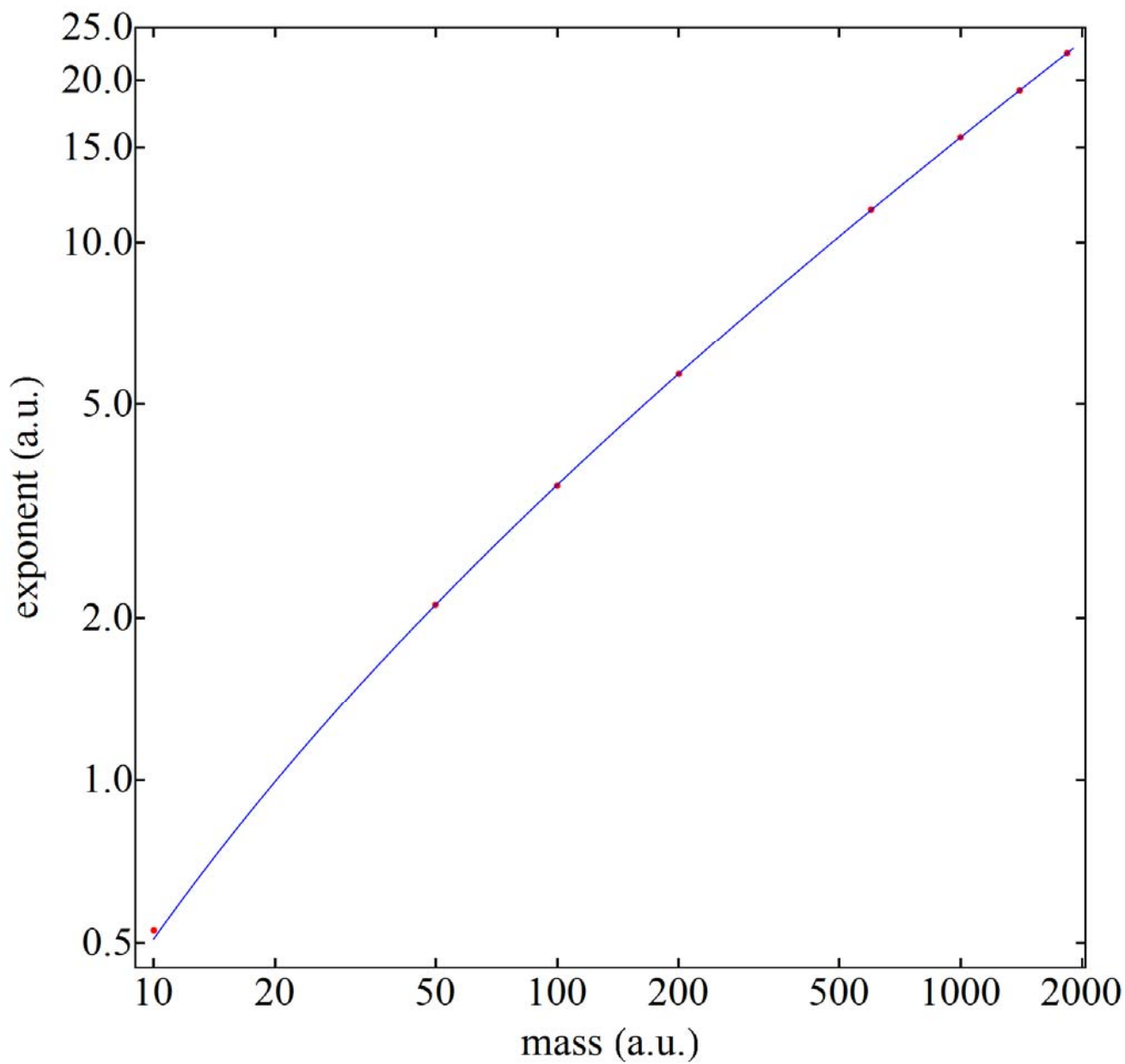

Figure S1- The logarithmic plot of the exponents of nuclear Gaussian basis functions versus the mass of corresponding positively charged particles using the data from the Hartree product wavefunction. The red dots are ab initio data in the range of $m_+ = 10m_e - 2000m_e$ while the blue curve is the equation emerging from regression procedure. The blue curve is based on following equation: $\alpha(m_+) = (a' + b'm_+^p)^2$ where $a' = -0.787722$, $b' = 0.843870$, $p = 0.250017$.

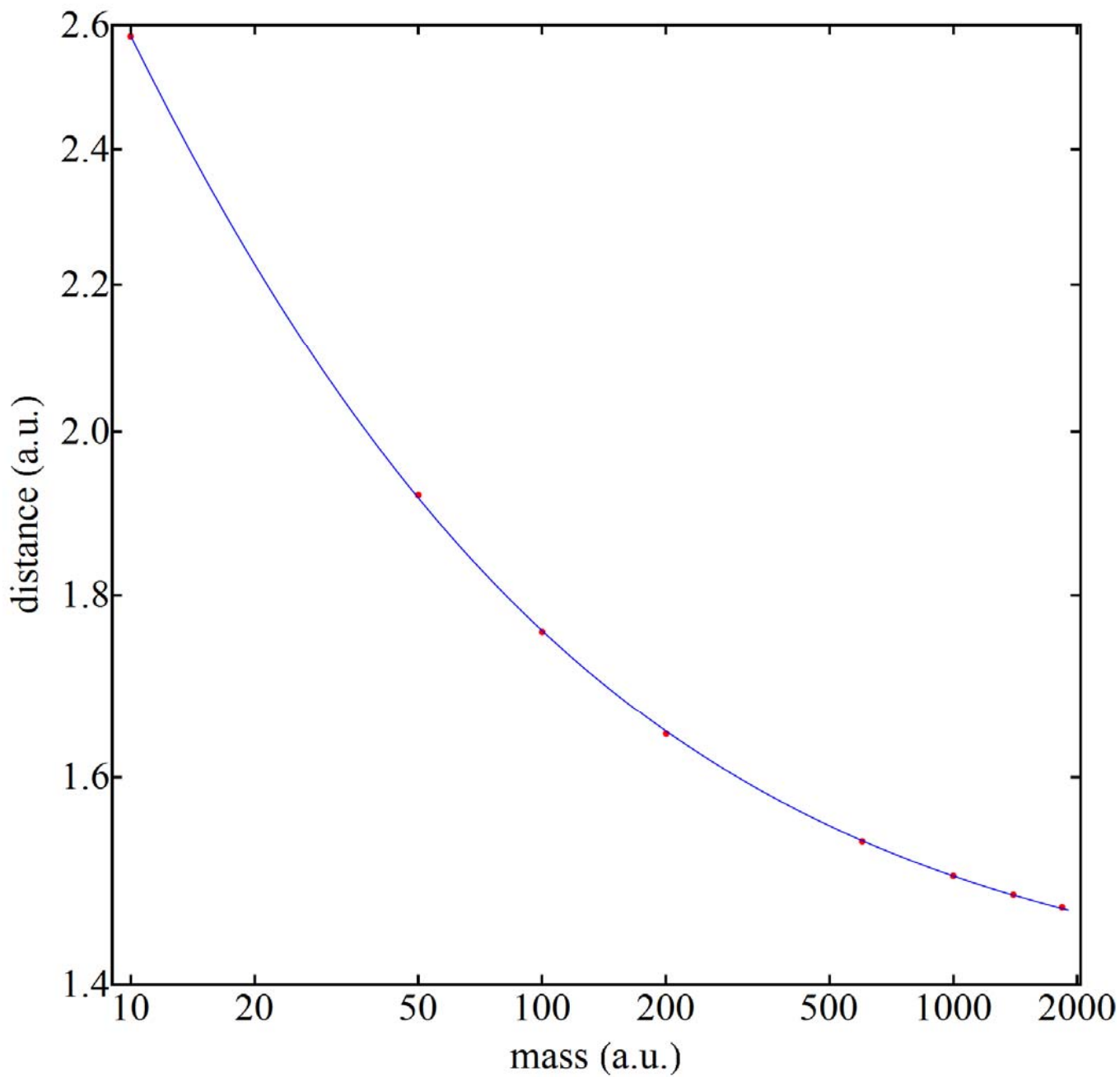

Figure S2- The logarithmic plot of the distance between nuclear Gaussian basis functions versus the mass of corresponding positively charged particles using the data from the Hartree product wavefunction. The red dots are ab initio data in the range of $m_+ = 10m_e - 2000m_e$ while the blue curve is the equation emerging from regression procedure. The blue curve is based on following equation: $d(m_+) = am_+^{-n} + b$ where $a = 3.819156$, $b = 1.383523$, $n = 0.503466$.

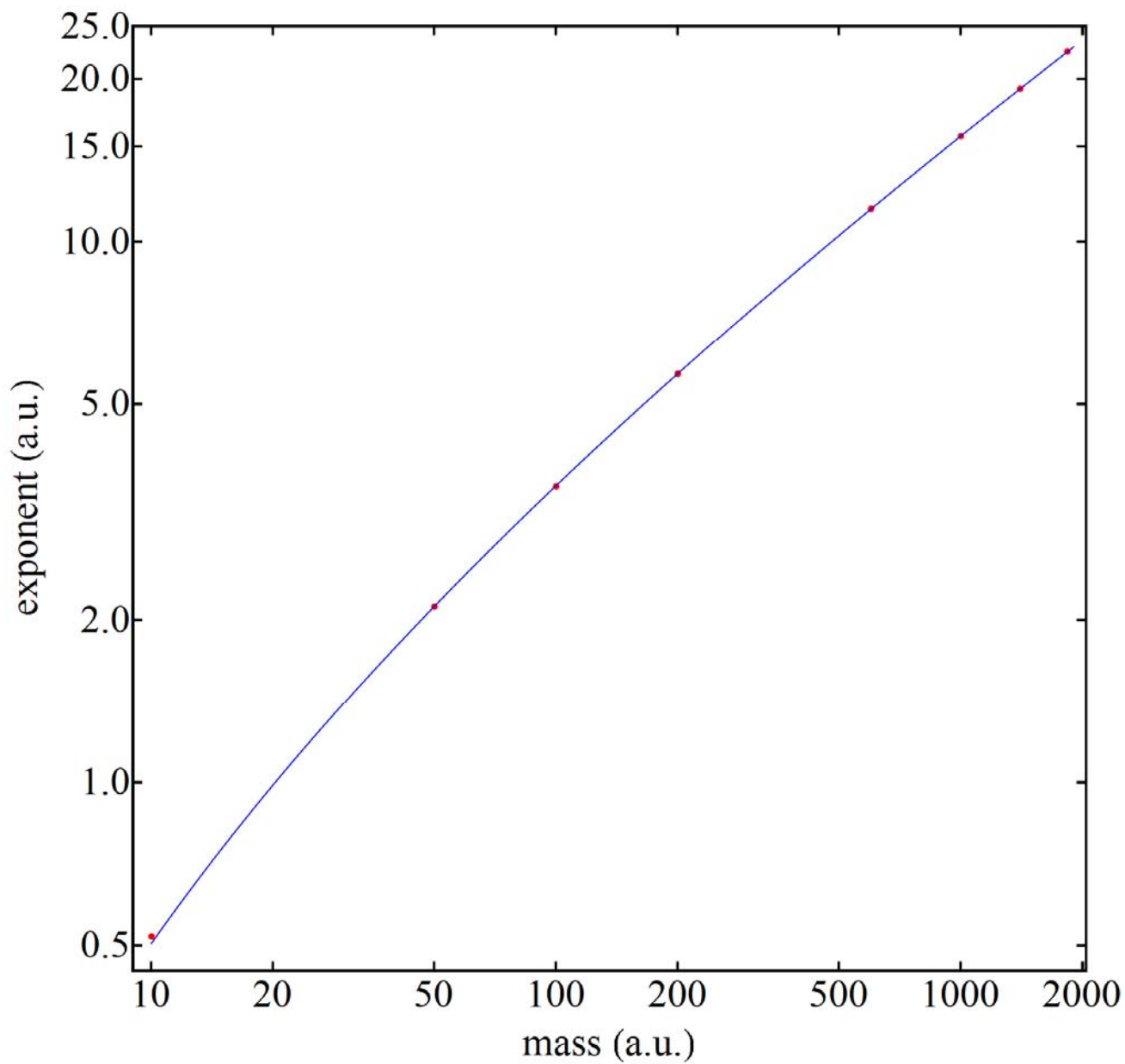

Figure S3- The logarithmic plot of the exponents of nuclear Gaussian basis functions versus the mass of corresponding positively charged particles using the data from the Hartree-Fock singlet/para wavefunction. The red dots are ab initio data in the range of $m_+ = 10m_e - 2000m_e$ while the blue curve is the equation emerging from regression procedure. The blue curve is based on following equation: $\alpha(m_+) = (a' + b'm_+^p)^2$ where $a' = -0.796678$, $b' = 0.848673$, $p = 0.249466$.

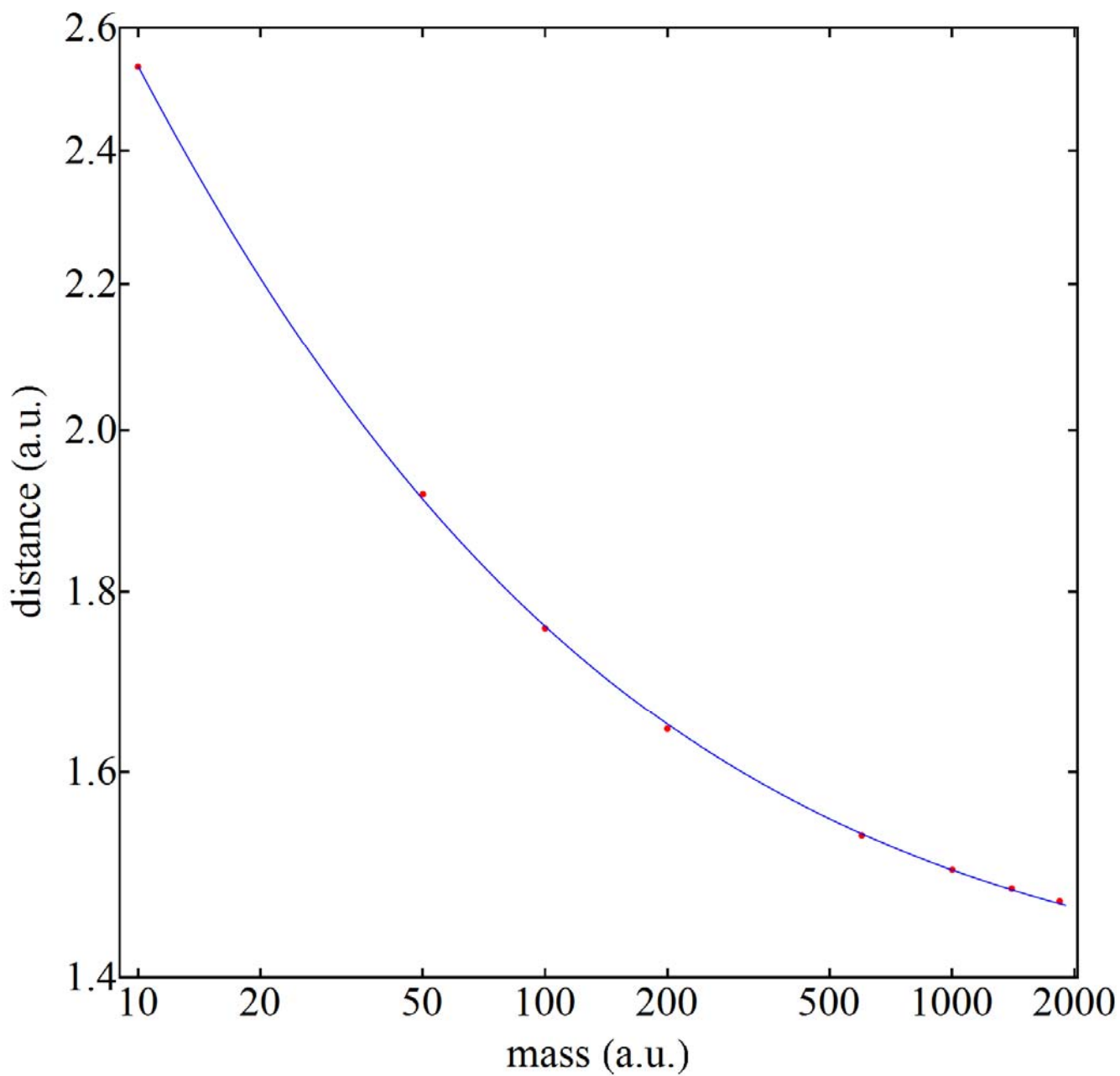

Figure S4- The logarithmic plot of the distance between nuclear Gaussian basis functions versus the mass of corresponding positively charged particles using the data from the Hartree-Fock singlet/para wavefunction. The red dots are ab initio data in the range of $m_+ = 10m_e - 2000m_e$ while the blue curve is the equation emerging from regression procedure. The blue curve is based on following equation: $d(m_+) = am_+^{-n} + b$ where $a = 3.489295$, $b = 1.372144$, $n = 0.476910$.

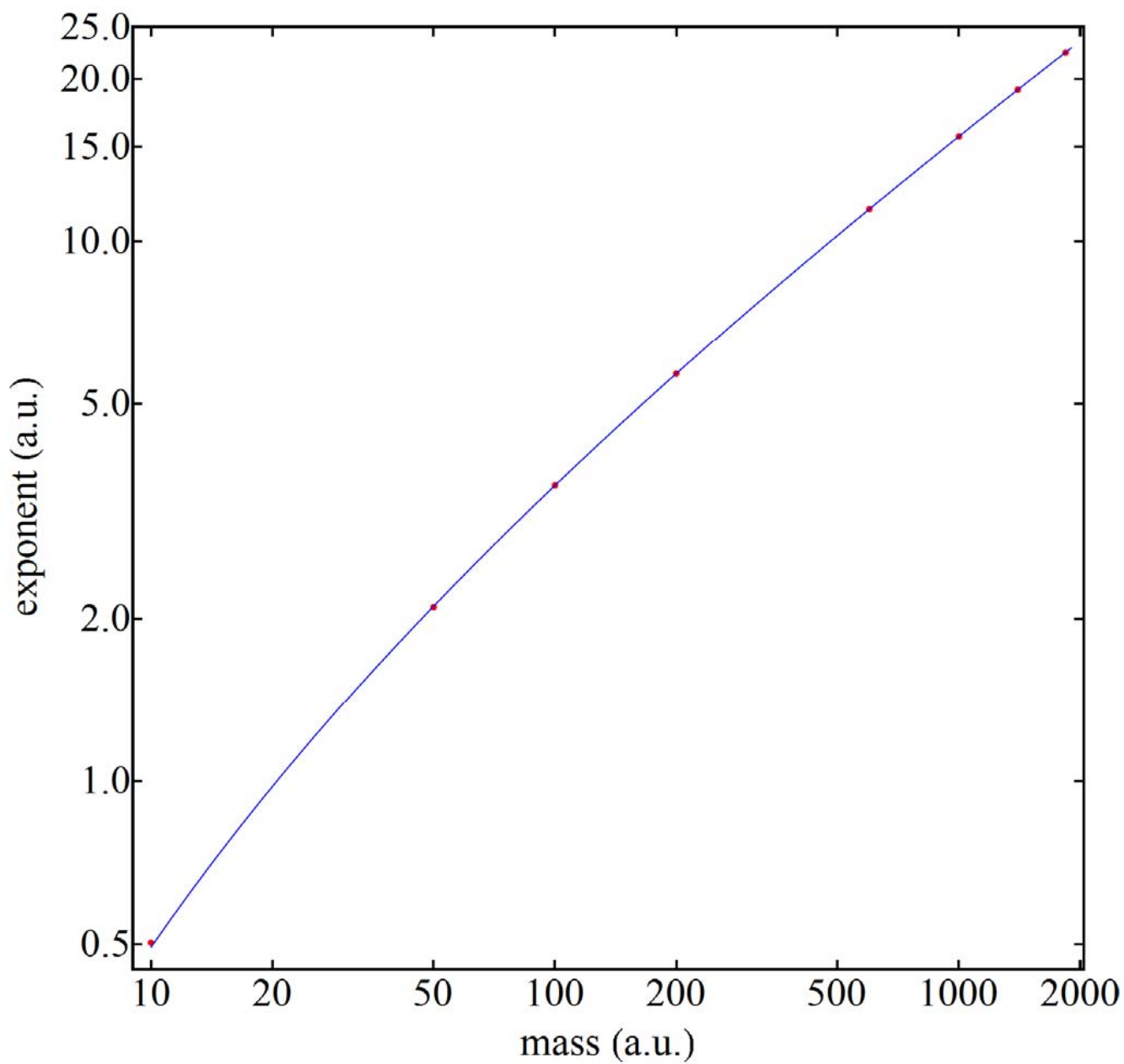

Figure S5- The logarithmic plot of the exponents of nuclear Gaussian basis functions versus the mass of corresponding positively charged particles using the data from the Hartree-Fock triplet/ortho wavefunction. The red dots are ab initio data in the range of $m_+ = 10 m_e - 2000 m_e$ while the blue curve is the equation emerging from regression procedure. The blue curve is based on following equation: $\alpha(m_+) = (a' + b' m_+^p)^2$ where $a' = -0.823951$, $b' = 0.862790$, $p = 0.247918$.

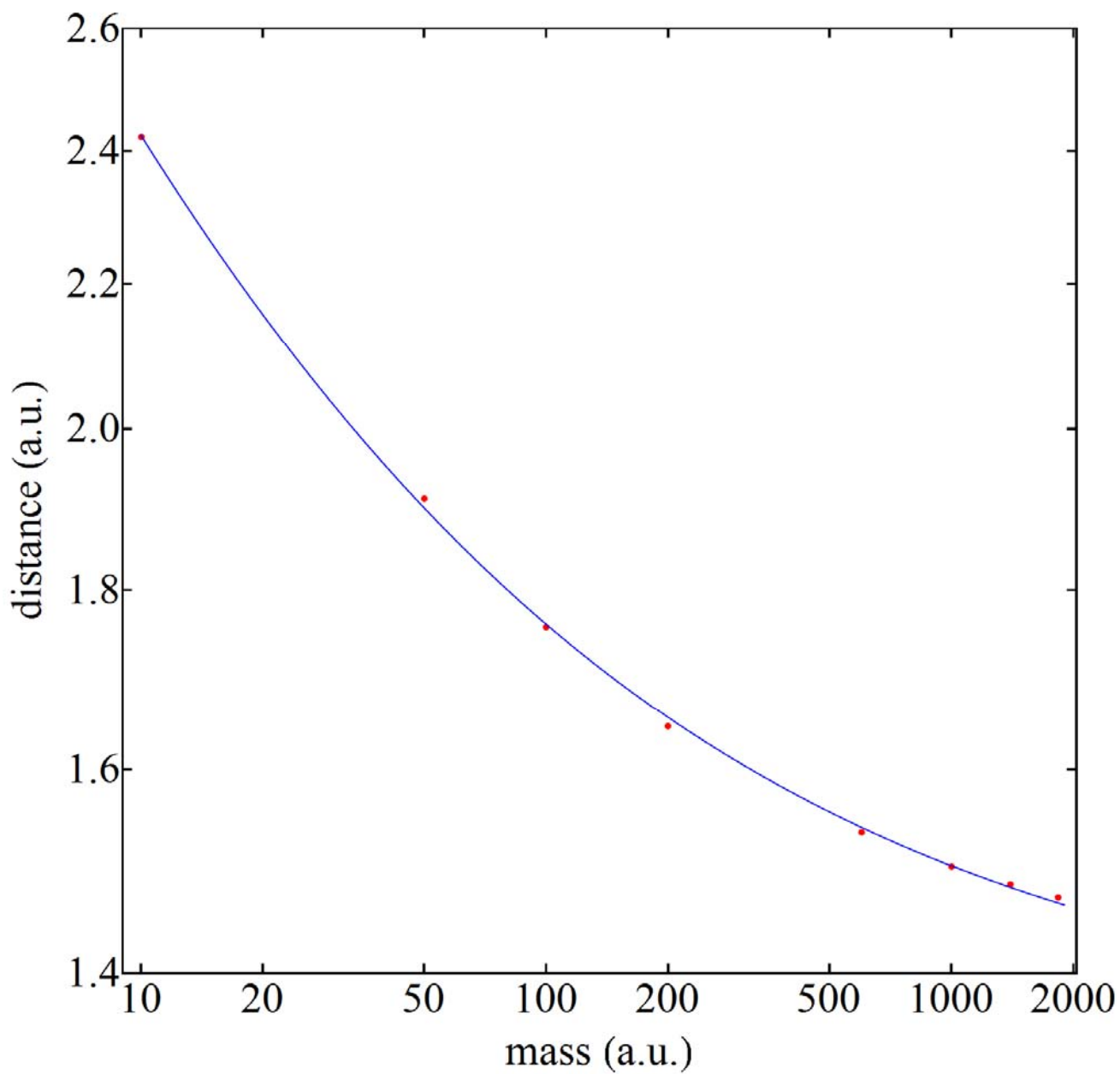

Figure S6- The logarithmic plot of the distance between nuclear Gaussian basis functions versus the mass of corresponding positively charged particles using the data from the Hartree-Fock triplet/ortho wavefunction. The red dots are ab initio data in the range of $m_+ = 10 m_e - 2000 m_e$ while the blue curve is the equation emerging from regression procedure. The blue curve is based on following equation: $d(m_+) = a m_+^{-n} + b$ where $a = 2.795882$, $b = 1.338276$, $n = 0.410672$.

Table S1- The components of the topological indexes calculated at the (3, -3) CPs of $\Gamma^{(2)}(\vec{q})$ employing the HP wavefunction with [6s:1s] basis set all offered in atomic units.

| mass | $G_+$ | $G_-$ | $\tilde{G}$ | $K_+$ | $K_-$ | $\tilde{K}$ | $\nabla^2 \rho_+/m$ | $\nabla^2 \rho_-$ | $\nabla^2 \Gamma^{(2)}$ |
|---|---|---|---|---|---|---|---|---|---|
| 10 | 0.0019 | 0.0004 | 0.0023 | 0.0242 | 0.0344 | 0.0585 | -0.0892 | -0.1357 | -0.2249 |
| 50 | 0.0073 | 0.0016 | 0.0089 | 0.1680 | 0.2028 | 0.3708 | -0.6427 | -0.8047 | -1.4474 |
| 100 | 0.0123 | 0.0023 | 0.0146 | 0.3053 | 0.3596 | 0.6649 | -1.1720 | -1.4292 | -2.6012 |
| 200 | 0.0191 | 0.0030 | 0.0220 | 0.5080 | 0.5824 | 1.0905 | -1.9559 | -2.3179 | -4.2738 |
| 600 | 0.0338 | 0.0040 | 0.0378 | 0.9964 | 1.0883 | 2.0848 | -3.8504 | -4.3374 | -8.1878 |
| 1000 | 0.0427 | 0.0044 | 0.0471 | 1.3050 | 1.3904 | 2.6954 | -5.0491 | -5.5439 | -10.5930 |
| 1400 | 0.0462 | 0.0044 | 0.0506 | 1.5552 | 1.7109 | 3.2661 | -6.0360 | -6.8260 | -12.8620 |
| H | 0.0553 | 0.0049 | 0.0602 | 1.7481 | 1.8028 | 3.5510 | -6.7714 | -7.1916 | -13.9629 |
| D | 0.0675 | 0.0051 | 0.0726 | 2.3941 | 2.5194 | 4.9135 | -9.3063 | -10.0574 | -19.3637 |
| T | 0.0819 | 0.0056 | 0.0875 | 2.7999 | 2.6757 | 5.4756 | -10.8718 | -10.6805 | -21.5523 |
| $10^4$ | 0.1077 | 0.0063 | 0.1140 | 3.4903 | 3.1523 | 6.6427 | -13.5306 | -12.5841 | -26.1147 |
| $10^5$ | 0.2507 | 0.0023 | 0.2530 | 7.4165 | 5.2070 | 12.6236 | -28.6633 | -20.7950 | -49.4584 |
| $10^6$ | 0.5153 | 0.0095 | 0.5247 | 14.5474 | 9.2071 | 23.7545 | -56.1284 | -36.7905 | -92.9190 |
| $10^7$ | 1.2273 | 0.0123 | 1.2396 | 25.7470 | 11.2513 | 36.9982 | -98.0787 | -44.9558 | -143.0346 |
| $10^8$ | 0.8674 | 0.0000 | 0.8674 | -0.3004 | 11.7694 | 11.4690 | 4.6710 | -47.0775 | -42.4065 |
| $10^9$ | 0.0005 | 0.0000 | 0.0005 | -0.0004 | 12.0999 | 12.0995 | 0.0033 | -48.3997 | -48.3963 |
| $10^{10}$ | 0.0000 | 0.0000 | 0.0000 | 0.0000 | 12.2120 | 12.2120 | 0.0000 | -48.8480 | -48.8480 |
| $10^{11}$ | 0.0000 | 0.0000 | 0.0000 | 0.0000 | 12.1565 | 12.1565 | 0.0000 | -48.6262 | -48.6262 |
| $10^{12}$ | 0.0000 | 0.0000 | 0.0000 | 0.0000 | 12.2589 | 12.2589 | 0.0000 | -49.0358 | -49.0358 |
| $10^{13}$ | 0.0000 | 0.0000 | 0.0000 | 0.0000 | 12.2592 | 12.2592 | 0.0000 | -49.0369 | -49.0369 |